\pdfoutput=1
\documentclass[pre,aps]{revtex4}

\usepackage{times}
\usepackage{amsmath}
\usepackage{graphicx}
\usepackage{amssymb}
\usepackage{color}

\newcommand{\ii}{\mathrm{i}}

\newcommand{\cP}{\ensuremath{\mathcal{P}}}
\newcommand{\cT}{\ensuremath{\mathcal{T}}}

\newcommand{\ep}{\epsilon}
\newcommand{\p}{\partial}

\begin{document}

\title{$\cP\cT$-symmetric sine-Gordon breathers}

\author{Nan Lu}
\affiliation{Department of Mathematics and Statistics, University of
Massachusetts, Amherst, MA 01003-9305, USA}

\author{Jes\'us Cuevas-Maraver}
\affiliation{Grupo de F\'{\i}sica No Lineal, Departamento de F\'{\i}sica Aplicada I, Universidad de Sevilla. Escuela Polit{\'e}cnica Superior, C/ Virgen de \'Africa, 7, 41011-Sevilla, Spain. \\
Instituto de Matem\'aticas de la Universidad de Sevilla (IMUS).
Edificio Celestino Mutis. Avda. Reina Mercedes s/n, 41012-Sevilla, Spain
}

\author{Panayotis\ G.\ Kevrekidis}
\affiliation{Department of Mathematics and Statistics, University of
Massachusetts, Amherst, MA 01003-9305, USA}


\begin{abstract}
In this work, we explore a prototypical example of a genuine continuum
breather (i.e., not a standing wave) and the conditions under which it
can persist in a $\mathcal{P T}$-symmetric medium. As our model of interest,
we will explore the sine-Gordon equation in the presence of a $\mathcal{P T}$-
symmetric perturbation. Our main finding is that the breather of the
sine-Gordon model will {\it only} persist at the interface between
gain and loss that $\mathcal{P T}$-symmetry imposes but will not be
preserved if centered at the lossy or at the gain side. The latter
dynamics is found to be interesting in its own right giving rise
to kink-antikink pairs on the gain side and complete decay of the breather
on the lossy
side. Lastly, the stability of the breathers centered at the interface
is studied. As may be anticipated on the basis of their ``delicate''
existence properties such breathers are found to be destabilized
through
a Hopf bifurcation in the corresponding Floquet analysis.
\end{abstract}
\maketitle

\section{Introduction}

Over the past 15 years i.e., after its theoretical suggestion in
the realm of linear quantum mechanics~\cite{bend1,bend2,bend3},
the examination of systems combining gain and loss in a so-called
$\mathcal{P T}$-symmetric form has seen an explosive increase of
interest. $\mathcal{P T}$-symmetry implies that the gain and loss
are introduced in a way such that the system remains invariant
under the combined transformation of $x \rightarrow -x$ and
$t \rightarrow -t$ (and $i \rightarrow -i$ in Schr{\"o}dinger type
settings of a complex order parameter). This physically implies
a balance of the gain and loss regions that leads to an intriguing
class of systems bearing both ``Hamiltonian'' characteristics
(such as mono-parametric families of solutions and symmetric
spectra) and ``dissipative'' ones (such as non-energy conserving
evolutionary dynamics).

One of the principal sources for the considerable interest in
this class of systems has been
the realization that areas such as optics might be well suited both
for the theoretical study (as per the pioneering suggestions
of~\cite{Muga,ziad,Ramezani}),
and for the experimental realization~\cite{salamo,dncnat,whisper}
of such systems. In fact, more recently additional areas of
application of $\mathcal{P T}$-symmetric systems have also emerged.
In particular, measurements of $\mathcal{P T}$-symmetric
realizations (and identifications e.g. of the so-called
$\mathcal{P T}$-phase transition) have taken place at the mechanical
level~\cite{bend_mech} and at the electrical one~\cite{R21,tsampas2}.
These experimental efforts have given rise to a large volume of
theoretical literature on the study of solitary waves,
breathers and even rogue waves in continua and lattices~\cite{Abdullaev,baras2,baras1,malom1,malom2,Nixon,Dmitriev,Pelin1,Sukh,bludov,yang2},
as well as on the complementary aspect of low-dimensional (oligomer or
plaquette) settings~\cite{Li,Guenter,suchkov,Sukhorukov,ZK,Flachbar,Flachbar2}.

Admittedly, most of these works have been focused on the Schr{\"o}dinger
class of dynamical models (of a complex order parameter) in which the
notion of $\mathcal{P T}$-symmetry was originally
proposed~\cite{bend1,bend2,bend3} (and
also experimentally implemented~\cite{salamo,dncnat}). Nevertheless,
recently there have been various motivations for exploring similar
notions in Klein-Gordon type systems.
At the experimental linear level, a relevant example is the dimer
configuration of coupled gain-loss second order oscillators
of~\cite{R21,tsampas2}. Recently, also,
a Klein-Gordon setting has also been explored theoretically
for so-called $\mathcal{P T}$-symmetric nonlinear metamaterials and
the formation of gain-driven discrete breathers therein~\cite{tsironis}.
This, in turn, has led to a number of studies of discrete~\cite{ptkg,demirkaya}
(both at the level of oligomers~\cite{ptkg} and at that of
lattices~\cite{demirkaya}) and continuum~\cite{demirkaya2}
Klein-Gordon models of $\mathcal{P T}$-symmetric media.
It is relevant to point out that nonlinear dimers of
along the lines of second order evolution equations are progressively
attracting more attention both at the theoretical~\cite{gianfreda}
and even at the experimental~\cite{factor} level.

While the studies in the case of extended systems
have explored predominantly the realm of standing waves (in nonlinear
Schr{\"o}dinger systems) and kinks (in Klein-Gordon ones), genuine
Klein-Gordon breather
states have not been addressed (to the best of our knowledge). The latter,
as is well known, are far more delicate~\cite{birnir}, especially so
in the continuum limit when subject to perturbations. It is the aim
of the present study to indeed explore the existence, stability
and dynamics of continuum breathers in, arguably, the prototypical
model in which they exist namely the sine-Gordon equation under
$\mathcal{P T}$-symmetric perturbations. We theoretically derive
conditions for the persistence of the breathers, which illustrate
that they can survive at the interface between the region of gain
and that of loss within our $\mathcal{P T}$-symmetric medium (section II).
We then numerically explore their linear stability by means of
Floquet analysis and identify their instability (section III).
In the same section, we numerically explore their evolution
dynamics identifying distinct scenarios when the instability pushes
the breather on the gain side, vs. that of the loss side. In the former
case, a kink-antikink pair nucleates, while in the latter the breather
is found to be annihilated. Lastly, in section IV, we summarize our findings
and present some potential directions for future study.

\section{Theoretical Analysis}

We consider a modified sine-Gordon equation of the form:

\begin{eqnarray}\label{dyn}
    u_{tt}-u_{xx} +\epsilon \gamma(x) u_t + \sin u = 0.
\end{eqnarray}

Here, in order to preserve the $\mathcal{P T}$-symmetry, $\gamma(x)$, should be an antisymmetric function satisfying $\gamma(-x)=-\gamma(x)$. This physically implies that while this is an ``open'' system with gain and loss,
where the gain balances the loss, in preserving the symmetry. In order to study the $\mathcal{PT}$ symmetry effects on breathers within a concrete example,
we have chosen $\gamma(x)=x\exp[-(x/2)^2]$.

When $\ep=0$, Eq.~\eqref{dyn} has even and odd in time breathers of the form:
\begin{equation}\label{eq1.1}
b_e(x,t)=4\arctan{\frac{\sigma\cos {at}}{a\cosh{\sigma x}}}\ \ \text{and}\ \ b_o(x,t)=4\arctan{\frac{\sigma\sin {at}}{a\cosh{\sigma x}}}\end{equation}
where $0<a<1$ and $\sigma=\sqrt{1-a^2}$; in what follows, however,
we will restrict our analytical considerations to breathers
with $a>1/2$ for technical reasons, briefly explained below.
More generally, these solutions are both space and time translation
invariant (associated, respectively, with momentum and energy conservation).
Yet, notice that here we only identify and utilize the even and odd
(in $t$) elements of the time translation invariant family of breather
solutions. Additionally, we will only be concerned about spatial
translations at a later, separate stage below.
The breather solutions can be obtained by the inverse scattering method. Theoretical studies of persistence of breathers under Hamiltonian perturbations can be found in \cite{birnir,denzler}, where the authors examined the rigidity of the breather. More precisely, they showed the breather can only survive under perturbations in a very specific form. Furthermore, the perturbed breather is just a rescaling of the unperturbed one. In general, the breather should deform to a family of solutions with small oscillations at infinity along the $x$-direction \cite{shatah,lu}.

One way to study the persistence of breathers is to use spatial dynamics, namely, swapping $x$ and $t$. Under such a formulation, breathers can be viewed as homoclinic orbits for a nonlinear wave equation with periodic boundary conditions. Therefore, invariant manifold theory and Melnikov type analysis can be used to establish the persistence of breathers. We will adopt such a formulation in
what follows.
Consider the equation (obtained by swapping $x$ and $t$ in \eqref{dyn})
\begin{equation}\label{eq2}
u_{tt}=(1+a^2\p_{xx}) u+\ep a \gamma(t) u_x+(\sin u-u).
\end{equation}
The parameter $a$ comes from rescaling so that we can consider \eqref{eq2} under $2\pi$-periodic boundary condition for every $a$. The breathers then
become (within spatial dynamics)
\begin{equation}\label{eq2.1}
b_e(x,t)=4\arctan{\frac{\sigma\cos {x}}{a\cosh{\sigma t}}}\ \ \text{and}\ \ b_o(x,t)=4\arctan{\frac{\sigma\sin {x}}{a\cosh{\sigma t}}},
\end{equation}
where the subscripts $e,o$ refer to the breather that is even or odd in $x$.
The linear operator $1+a^2\p_{xx}$ has characteristic
frequencies $\sqrt{1-a^2k^2}$, where $k=0,1,\cdots$. Therefore, the
corresponding real eigenvalues $\pm 1$ have
multiplicity $1$ and $\pm\sigma$ (specifically under the choice
made above of $a>1/2$) have multiplicity $2$ (this is because we
cannot assume $u$ is even or odd in $x$ in \eqref{eq2}). All other
eigenvalues are $\pm i\sqrt{a^2k^2-1}$ with multiplicity $2$.
From the distribution of eigenvalues (3-stable modes, 3-unstable modes,
infinitely many neutral modes), it is not hard to see why the breather is an
uncommon feature (at least for continuum models). For a
general Klein-Gordon type equation, one should expect to find solutions
that converge to oscillations generated by those neutral modes instead of
the stationary solution $0$. This simple observation conceptually confirms
the theoretical results  mentioned above.

If we now rewrite \eqref{eq2} as a first order system, we have
\begin{equation}\label{firstorder}
\left\{\begin{aligned}
&u_t=v,\\
&v_t=(1+a^2\p_{xx}) u+\ep a \gamma(t) u_x+(\sin u-u).
\end{aligned}\right.
\end{equation}
Recall that when $\ep=0$, Eq.~\eqref{eq2} has the Hamiltonian
\[H(u,u_t)=\int \left[\frac 12 u_t^2+\frac{a^2}{2}u_x^2+\cos u\right]\ dx.\]
To keep our exposition clean, we focus on the odd breather, namely, $b_o$ in the following analysis. The results for even breathers $b_e$
follow from an essentially identical analysis to the case of
$b_o$. Formally, the persistence condition for the breather
is given by the Melnikov integral which assumes the form~\cite{guggen}:
\begin{equation}\label{eq7}
\begin{aligned}
M(t_0)=&\int_{-\infty}^{+\infty}\int_{-\pi}^{\pi}\nabla H(b_{o},\p_t b_{o})\begin{bmatrix}0\\a\gamma (t+t_0)\p_xb_{o}\end{bmatrix}\ dxdt\\
=&a\int_{-\infty}^\infty \int_{-\pi}^{\pi}(\p_t b_{o}) (\p_x b_{o}) \gamma(t+t_0)\ dxdt.
\end{aligned}
\end{equation}
Here $b_{o}$ is defined as in \eqref{eq2.1}. Note that $(\p_tb_{o})(\p_x b_{o})$ are odd in $x$, which implies
\begin{equation}\label{fm}
M(t_0)\equiv0.
\end{equation}
Therefore, the unstable manifold and center-stable manifold split at most
by a distance $O(\ep^2)$, which means that the above Melnikov
function is not particularly useful in this case.\\

In fact, the splitting of the unstable manifold and the
center-stable manifold can be measured by $H(\phi^u(t_0,\ep),\phi_t^u(t_0,\ep))-H(\phi^{cs}(t_0,\ep),\phi_t^{cs}(t_0,\ep))$, where $(\phi^u,\phi_t^u)$ and $(\phi^{cs},\phi_t^{cs})$ are solutions
that stay on the perturbed unstable manifold and center-stable manifolds. Since
\[|\phi^{u,cs}(t_0,\ep)-b_{o}|+|\phi_t^{u,cs}(t_0,\ep)-\p_t b_{o}|=O(\ep),\]
one can easily derive
\[H(\phi^u(t_0,\ep),\phi_t^u(t_0,\ep))-H(\phi^{cs}(t_0,\ep),\phi_t^{cs}(t_0,\ep))=\ep M(t_0)+O(\ep^2).\]
We next calculate the leading order term in $O(\ep^2)$. Formally, we have
\[H(\phi^u(t_0,\ep),\phi_t^u(t_0,\ep))-H(\phi^{cs}(t_0,\ep),\phi_t^{cs}(t_0,\ep))=H(\phi^u(t_0,\ep),\phi_t^u(t_0,\ep))-H(0,0)-H(\phi^{cs}(t_0,\ep),\phi_t^{cs}(t_0,\ep))+H(0,0).\]
Let $\Phi(t,\ep;u(t_0,\ep),u_t(t_0,\ep))$ be the solution map of \eqref{firstorder} at time $t$ with initial data $(u(t_0,\ep),u_t(t_0,\ep))$ and $V(t,\ep,\Phi)$ be the right hand side of \eqref{firstorder}. Differentiating \eqref{eq2} with respect to $\ep$ and setting $v(\cdot)=\p_\ep u|_{\ep=0}(\cdot+t_0)$ (suppressing
the $x$ variable for reasons of compactness) yield the first variational equation (for $\ep$)
\begin{equation}\label{eq8}
v_{tt}=(a^2\p_{xx}+\cos b_{o}) v+a\gamma(t+t_0)\p_x b_{o}(t+t_0).
\end{equation}
Moreover, $v$ is periodic and even in $x$. If we choose $b_e$, then $v$ is is odd in $x$.
The first Melnikov function in \eqref{fm} suggests
\begin{equation}\label{eq8.0}
v(0)=v_t(0)=0.\end{equation}
The invariance of $H$ for \eqref{firstorder} with $\ep=0$ implies
\[\int_{-\pi}^{\pi} DH(\Phi(t,\ep;u(t_0,\ep),u_t(t_0,\ep)))V(t,0,\Phi(t,\ep;u(t_0,\ep),u_t(t_0,\ep)))=0.\]
Consequently, we have
\begin{equation}\label{eq8.1}
\begin{aligned}
&H(\phi^u(t_0,\ep),\phi_t^u(t_0,\ep))-H(0,0)\\
=&\int_{-\infty}^0\int_{-\pi}^{\pi}\p_t H(\Phi(t,\ep;u(t_0,\ep),u_t(t_0,\ep)))\ dxdt\\
=&\int_{-\infty}^0\int_{-\pi}^{\pi} DH(\Phi(t,\ep;u(t_0,\ep),u_t(t_0,\ep)))\cdot \p_t\Phi(t,\ep;u(t_0,\ep),u_t(t_0,\ep))\ dxdt\\
=&\int_{-\infty}^0\int_{-\pi}^{\pi} DH(\Phi(t,\ep;u(t_0,\ep),u_t(t_0,\ep)))\cdot \Big(V(t,\ep,\Phi)-V(t,0,\Phi)\Big)\ dxdt\\
=&\int_{-\infty}^0\int_{-\pi}^{\pi} \Big(DH(b_o,\p_t b_o)+\ep D^2H(b_o,\p_t b_o)\cdot\frac{d}{d\ep}\Phi\Big|_{\ep=0}+O(\ep^2)\Big)\cdot \Big(\ep\int_0^1\p_\ep V(t,s\ep,\Phi)\ ds\Big)\ dxdt,
\end{aligned}\end{equation}
where $D^2H(b_{o},\p_t b_{o})$ is the Hessian of $H$ evaluated at $(b_{o},\p_t b_{o})$. Note that
\begin{equation}\label{eq8.2}
\frac{d}{d\ep}\Phi\Big|_{\ep=0}=\p_\ep\Phi(t,0;u(t_0,0),u_t(t_0,0))+D\Phi(t,0;u(t_0,0),u_t(t_0,0))\cdot\begin{bmatrix}v(t_0),v_t(t_0)\end{bmatrix}
=\begin{bmatrix}v(t),v_t(t)\end{bmatrix}.
\end{equation}
Here $D\Phi$ is the linearization of $\Phi$ with respect to initial data. The last equality follows from the definition of $v$ and
\eqref{eq8.0}. From the definition of $V$,
\begin{equation}\label{eq8.3}
\int_0^1\p_\ep V(t,s\ep,\Phi)\ ds=\begin{bmatrix}
0\\ a\gamma(t+t_0)(\p_xb(t+t_0)+\ep \p_x v+O(\ep^2))
\end{bmatrix}.
\end{equation}
Combining \eqref{eq8.1}, \eqref{eq8.2} and \eqref{eq8.3}, we have
\[\begin{aligned}
&H(\phi^u(t_0,\ep),\phi_t^u(t_0,\ep))-H(0,0)\\
=&-a \ep\int_{-\infty}^0\int_{-\pi}^{\pi}\p_t b_o(t+t_0)\p_x b_0(t+t_0)\gamma(t+t_0)\ dxdt\\
&-a\ep^2\int_{-\infty}^{0}\int_{-\pi}^{\pi} \Big(D^2H(b_{o},\p_t b_{o})\begin{bmatrix}v(t)\\v_t(t)\end{bmatrix}\Big)\cdot\begin{bmatrix}0\\ \gamma (t+t_0)\p_xb_{o}(t+t_0)\end{bmatrix} dxdt\\
&-a\ep^2\int_{-\infty}^{0}\int_{-\pi}^{\pi} \p_t b_{o}(t+t_0)\p_x v(t)\gamma(t+t_0)\ dxdt+O(\ep^3).
\end{aligned}\]
Similarly, one can derive
\[\begin{aligned}
&H(\phi^{cs}(t_0,\ep),\phi_t^u(t_0,\ep))-H(0,0)\\
=&a \ep\int_{+\infty}^0\int_{-\pi}^{\pi}\p_t b_o(t+t_0)\p_x b_0(t+t_0)\gamma(t+t_0)\ dxdt\\
&+a\ep^2\int_{+\infty}^{0}\int_{-\pi}^{\pi} \Big(D^2H(b_{o},\p_t b_{o})\begin{bmatrix}v(t)\\v_t(t)\end{bmatrix}\Big)\cdot\begin{bmatrix}0\\ \gamma (t+t_0)\p_xb_{o}(t+t_0)\end{bmatrix} dxdt\\
&+a\ep^2\int_{+\infty}^{0}\int_{-\pi}^{\pi} \p_t b_{o}(t+t_0)\p_x v(t)\gamma(t+t_0)\ dxdt+O(\ep^3).
\end{aligned}\]
Together with $M(t_0)=0$, we obtain
\[
\begin{aligned}
&H(\phi^u(t_0,\ep),\phi_t^u(t_0,\ep))-H(\phi^{cs}(t_0,\ep),\phi_t^{cs}(t_0,\ep))\\
=&-a\ep^2\int_{-\infty}^{+\infty}\int_{-\pi}^{\pi} \Big(D^2H(b_{o},\p_t b_{o})\begin{bmatrix}v(t)\\v_t(t)\end{bmatrix}\Big)\cdot\begin{bmatrix}0\\ \gamma (t+t_0)\p_xb_{o}(t+t_0)\end{bmatrix} dxdt\\
&-a\ep^2\int_{-\infty}^{+\infty}\int_{-\pi}^{\pi} \p_t b_{o}(t+t_0)\p_x v(t)\gamma(t+t_0)\ dxdt+O(\ep^3),
\end{aligned}
\]
The above integral can be simplified to (dropping the factor $-a\ep^2$)
\begin{equation}\label{eq9}
M_2(t_0)\triangleq\int_{-\infty}^{+\infty}\int_{-\pi}^{\pi}\big(\p_t b_{o}(t+t_0)\p_x v(t)+\p_t v(t)\p_x b_{o}(t+t_0)\big)\gamma(t+t_0)\ dxdt,
\end{equation}
where we recall that $b_{o}$ are defined in \eqref{eq2.1} and $v(t)$ satisfies \eqref{eq8}. For $t_0=0$, we observe that the solution of \eqref{eq8} is odd in $t$, because $\gamma$ is odd in $t$. Therefore, the integrand in \eqref{eq9} for $t_0=0$ is
\[(odd\times odd+even\times even)\times odd=odd\ in\ t,\]
which leads to
\[M_2(0)=0.\]
Next we calculate $\partial_{t_0} M_2(0)$, which satisfies
\[\begin{aligned}
\p_{t_0}M_2(0)
=&\int_{-\infty}^\infty\int_{-\pi}^\pi(\p_{tt} b_{o}\p_{x}v+\p_{t}v\p_{tx} b_{o})\gamma+(\p_t b_{o}\p_x v+\p_t v\p_x b_{o})\p_t\gamma\ dxdt\\
=&-\int_{-\infty}^\infty\int_{-\pi}^\pi(\p_{t}b_{o}\p_{tx} v+\p_{tt} v\p_{x}b_{o})\gamma\ dxdt,
\end{aligned}\]
where we integrate the second term in the first line by parts in $t$
to obtain the second line. Recall the fact from \eqref{eq8} that $v$ is even in $x$ for $b_o$. The integrand in the second line is
even both in $x$ and in $t$. Thus,
\[\p_{t_0}M_2(0)\ne0.\]
By going through a similar procedure, one can verify the same properties hold for $M_2(0)$ and $\partial_{t_0}M_2(0)$ if $b_o$ is replaced by $b_e$.

From the above spatial dynamics calculation, the result $M_2(0)=0$ implies
that the breather can be centered at the interface between gain and loss,
while the non-vanishing of the corresponding derivative suggests that the
relevant state does not persist (generically) either on the gain side
or on the lossy one. This confirms the delicate nature of this type of
coherent structure in the presence of $\mathcal{P T}$-symmetry. We now
turn to numerical computations for the stability and evolutionary dynamics
of the breather.

\section{Numerical Results}

In order to identify the relevant numerical breather solutions,
we have used a centered difference scheme to discretize the model
in space, while Fourier space techniques have been utilized in
order to expand the solution in time and
to obtain its numerically exact form (up to a prescribed
numerical tolerance). Finally, Floquet theory has been
used to explore the stability of the pertinent configurations.
More details about the numerical methods have been given in the Appendix.

We have studied the existence and stability of breathers centered at $x=0$ in a system that extends in the interval $[-50,50]$ starting from the Hamiltonian limit $\epsilon=0$ and continuing it to the $\cP\cT$ regime ($\epsilon>0$). To this aim, we have chosen a dissipation profile $\gamma(x)=x\exp[-(x/2)^2]$ and a breather frequency $a=0.9$. Notice that different frequencies give qualitatively similar results. In addition, we should indicate that using our Newton-Raphson type algorithm, we have confirmed that breathing solutions were indeed only tractable when centered at $x=0$, while our iteration failed to identify them for $x\neq0$.

\begin{figure}
\begin{tabular}{cc}
(a) & (b) \\
\includegraphics[width=6cm]{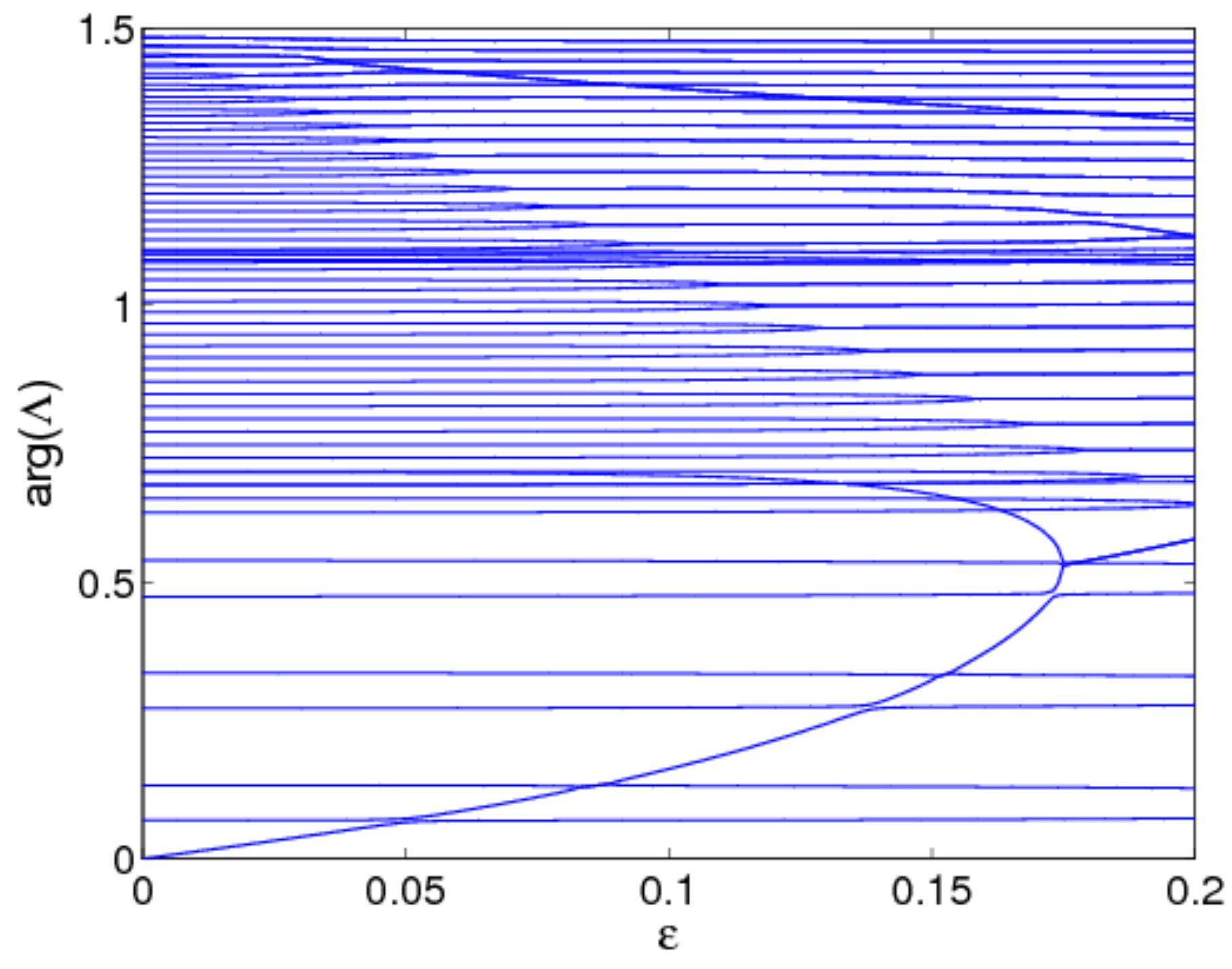} &
\includegraphics[width=6cm]{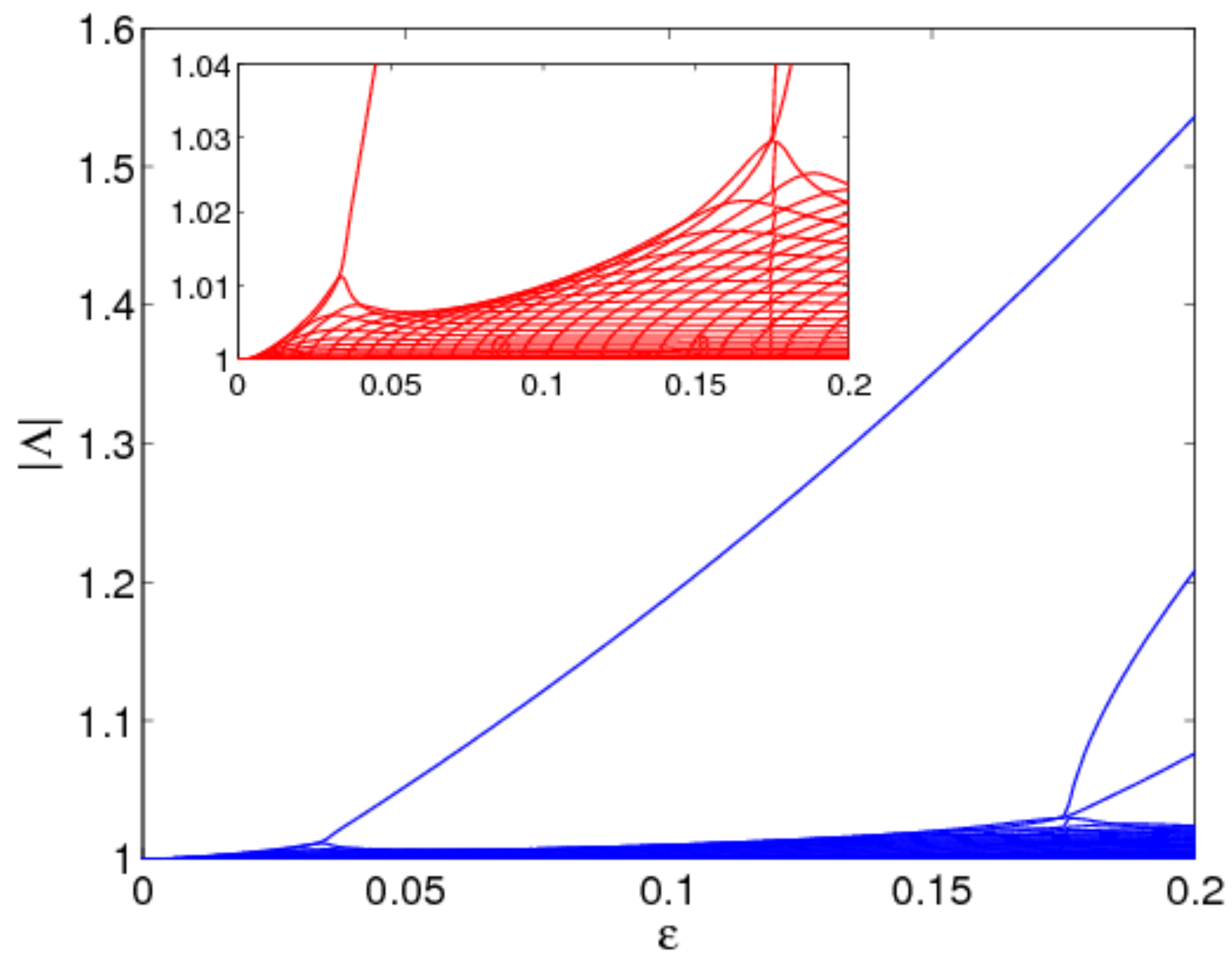} \\
\multicolumn{2}{c}{(c)} \\
\multicolumn{2}{c}{\includegraphics[width=6cm]{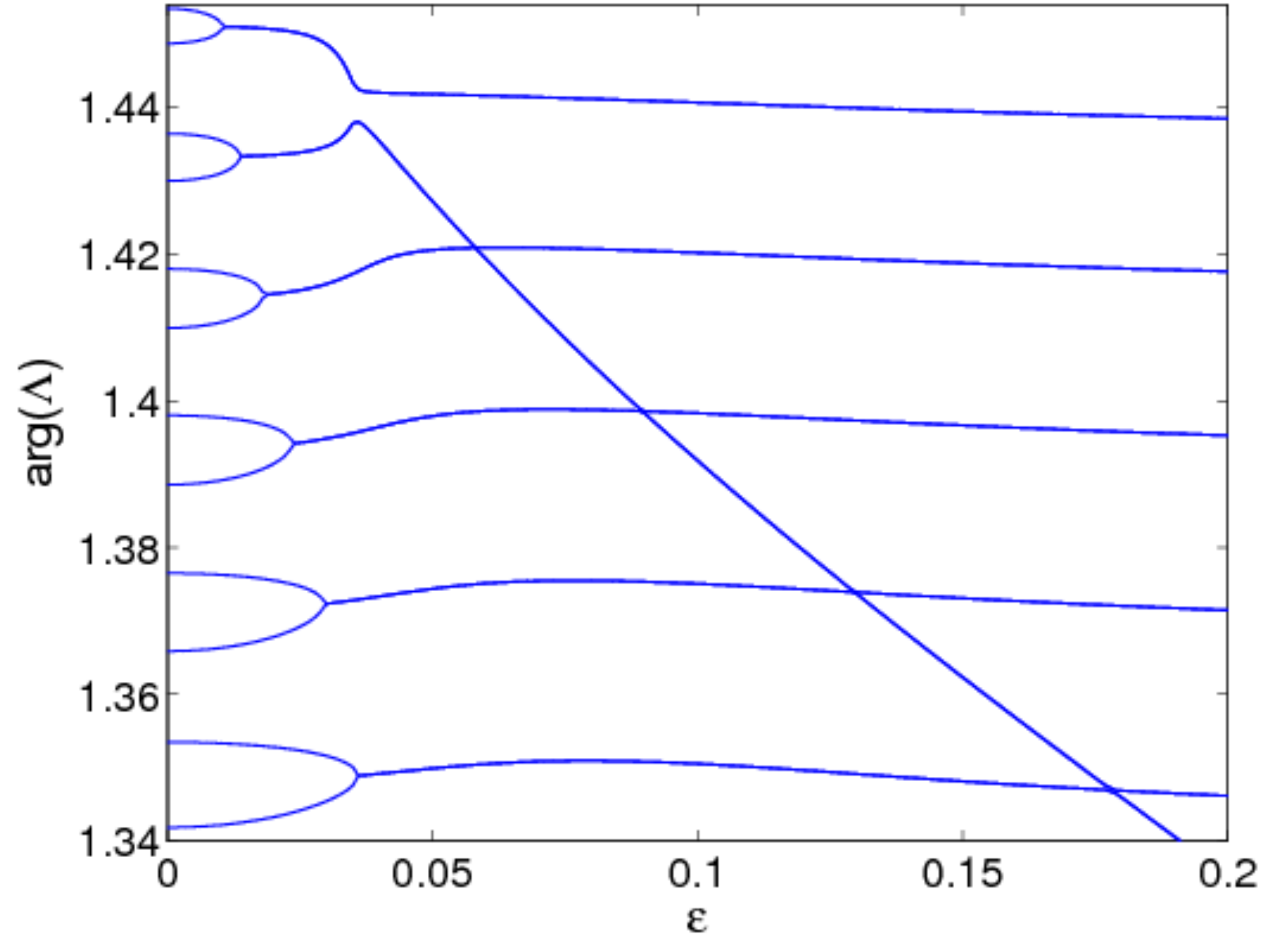}} \\
\\
\end{tabular}
\caption{Floquet spectrum versus $\epsilon$ for breathers with $a=0.9$ and discretization parameter $h=0.1$. Panel (a) shows the argument of the multipliers (i.e., their location on the trigonometric circle), while panel (b) shows their magnitude, which signals instability when it is $|\Lambda| > 1$ (notice that only $\theta\geq0$ and $|\Lambda|\geq1$ half-planes are shown, as the other half is found by symmetry).
It is worthwhile to point out the collisions around $\epsilon=0.014$ and $\epsilon=0.175$ in panel (a), which lead to the most significant instabilities in panel (b). Notice also that in panel (a), only the eigenvalues causing crossing or Hopf bifurcations are shown.
Panel (c) displays a zoom of panel (a) around the Hopf bifurcation causing the principal instabilities (i.e. highest moduli in panel b).}
\label{fig1}
\end{figure}

Next, we consider the stability properties of a breather centered at $x_0=0$. This breather, which is stable for every value of $a$ in the Hamiltonian
limit $\epsilon=0$ (see e.g. \cite{Aubry}) becomes unstable via a Hopf bifurcation when the gain/loss term is switched on
for a small value of $\epsilon>0$, as shown in Fig. \ref{fig1}.

The spectrum of the Floquet operator for the breather at the Hamitonian limit consists of two pairs of localized modes located at $+1$ and two symmetric (with respect to the real axis) bands of extended modes on the unit circle. The pairs of localized modes correspond to (1) the phase and growth modes and (2) the pinning (or translational) mode~\cite{flachgor}. The angles of the bands of extended modes can be approximated by $\theta\equiv\mathrm{arg}(\Lambda)\approx\pm2\pi\omega_\mathrm{ph}/a\ \mathrm{mod}\ 2\pi$ where $\omega_\mathrm{ph}$ are the frequencies of the linear modes (phonons) of the system (notice that the spectrum is wrapped around the unit circle). The above expression is approximate as the presence of the breather slightly deforms the (shape and frequency of the) modes and cause the appearance of a translational mode at $+1$. When the damping is switched on, translational invariance is broken, hence the translational mode departs from $\theta=0$; in addition, due to the localized character of the damping profile, extended modes start to become localized.

We find the breather to be unstable past $\epsilon_{c,1}=0.014$.
The arguments and magnitudes of different Floquet multipliers $\Lambda$
are shown in panels \ref{fig1}a and \ref{fig1}b, respectively.
Recall that instability is tantamount to $|\Lambda|>1$. Only eigenvalues causing crossing or bifurcations are shown in panel \ref{fig1}a for the sake of visibility. Notice that not every eigenvalue crossing is responsible for
collision, as the coincidence of two eigenvalue pairs
at a given $\theta$ is a necessary but not sufficient condition for the Hopf bifurcation to occur, even if the Krein signature of the coincident
eigenvalues is opposite, as shown by Aubry \cite{Aubry}. In addition, only eigenvalues with argument fulfilling $|\theta|<\pi/2$ are shown as the only observed collisions take place at this range.

\begin{figure}
\begin{tabular}{cc}
(a) & (b) \\
\includegraphics[width=6cm]{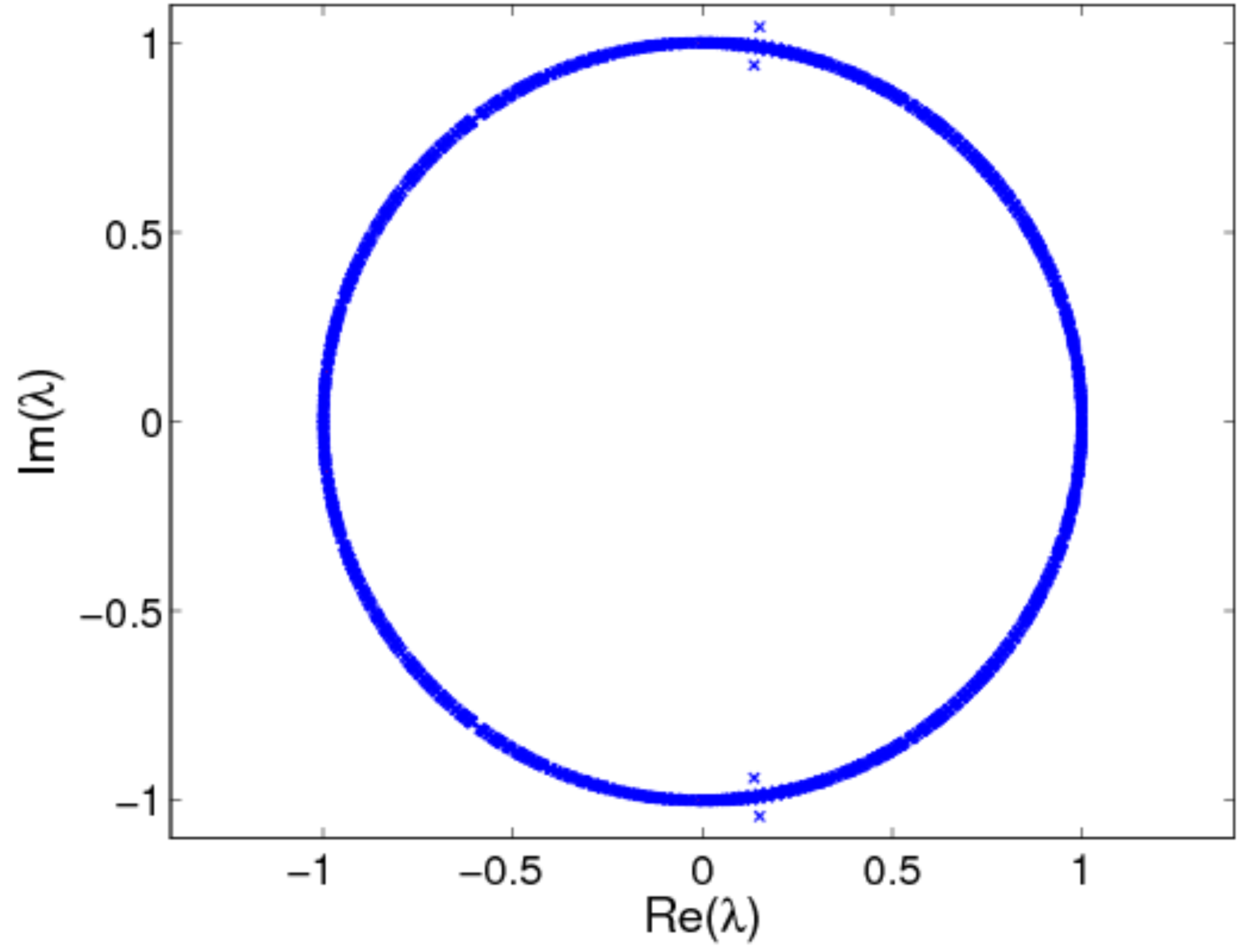} &
\includegraphics[width=6cm]{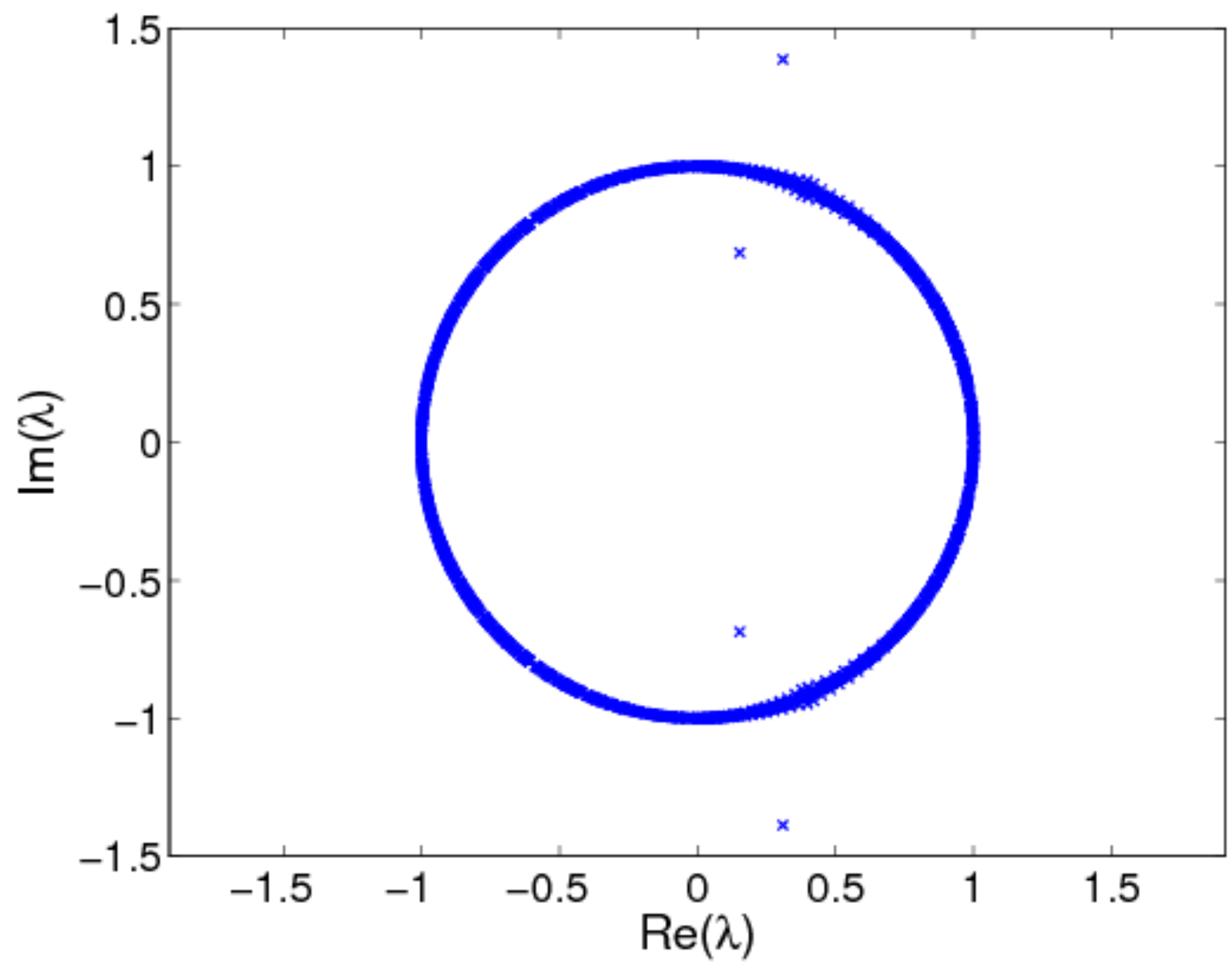} \\
(c) & (d) \\
\includegraphics[width=6cm]{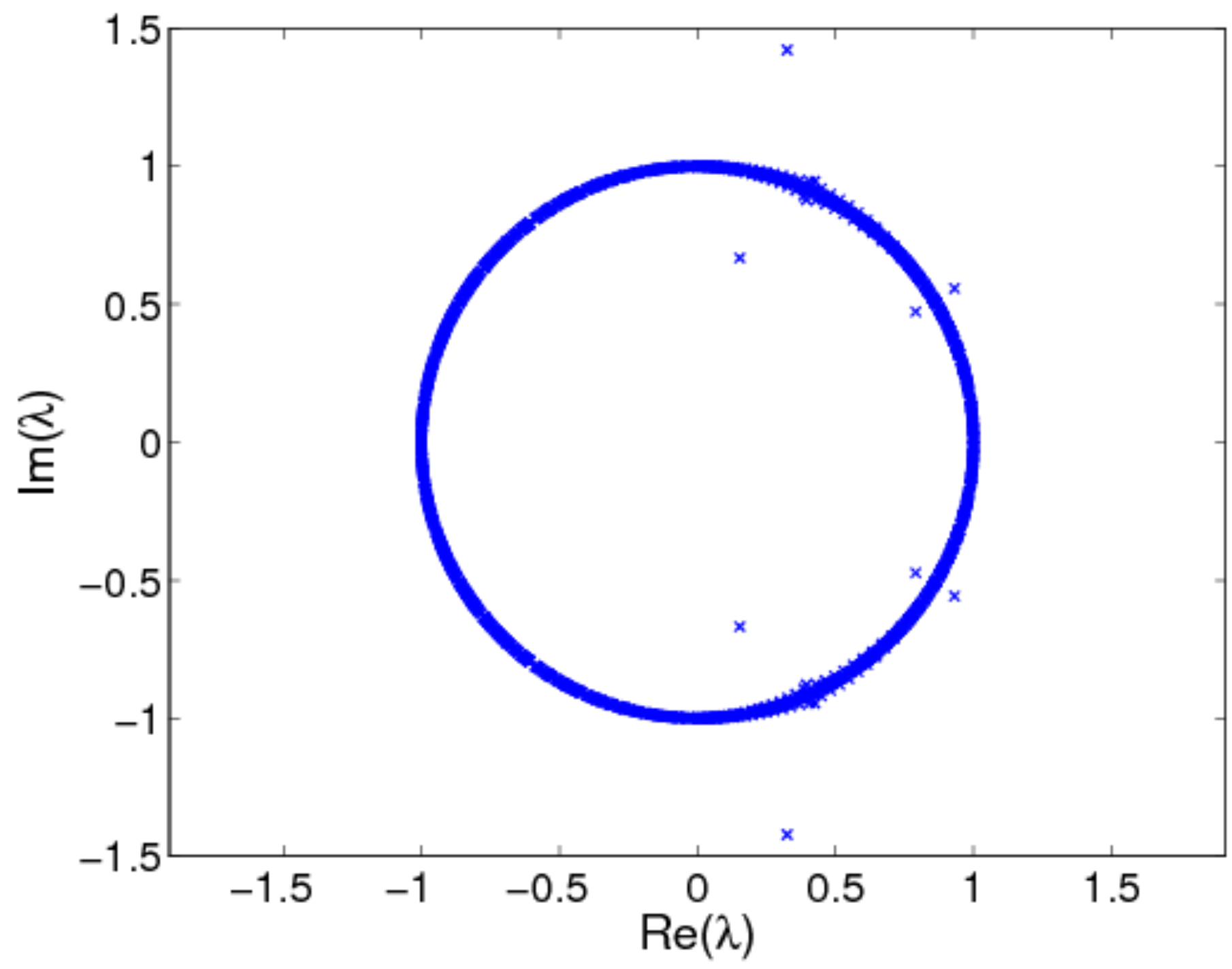} &
\includegraphics[width=6cm]{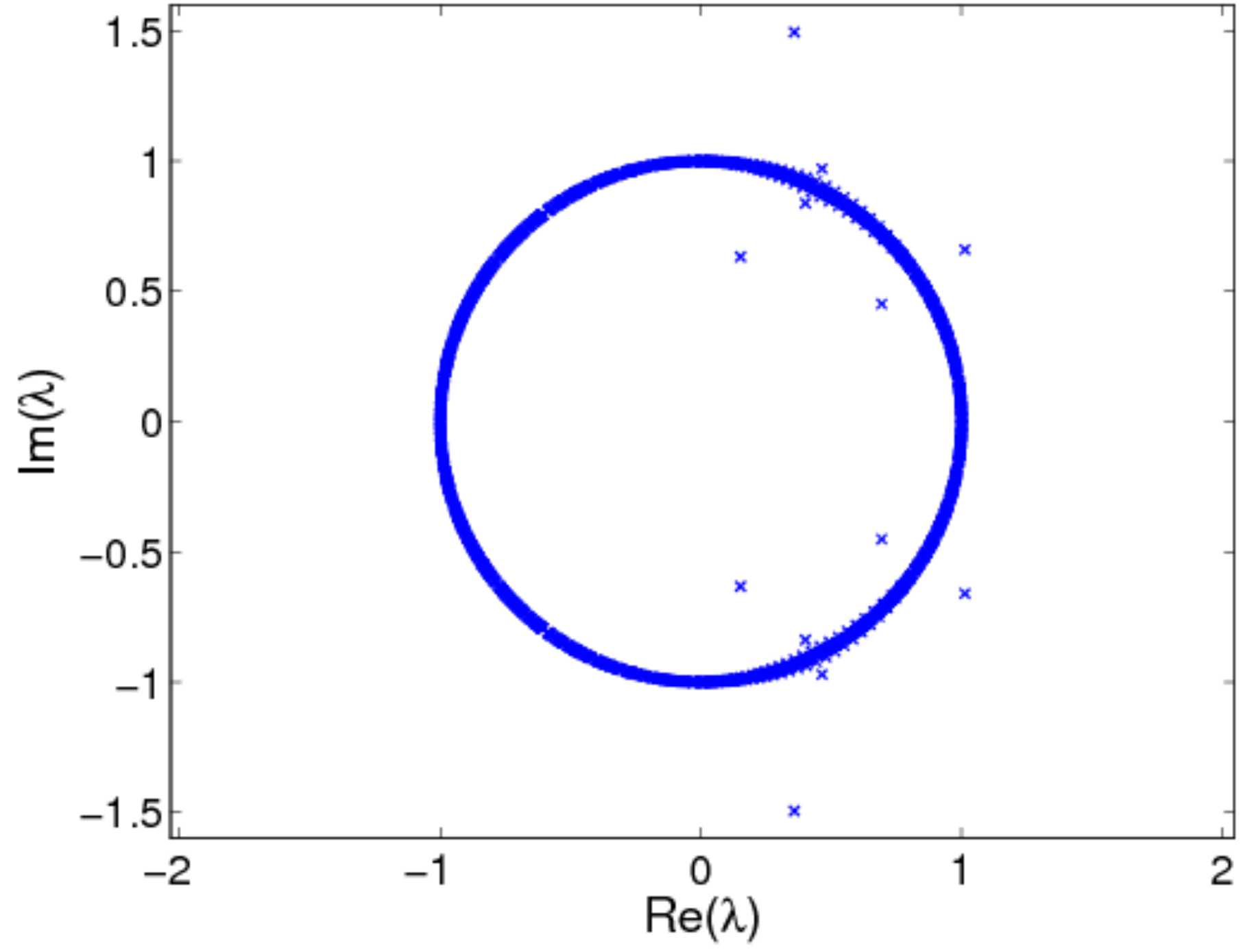} \\
\\
\end{tabular}
\caption{Full Floquet spectrum at (a) $\epsilon=0.05$, (b) $\epsilon=0.17$, (c) $\epsilon=0.18$ and (d) $\epsilon=0.2$. In the first case, only the instability caused by the bifurcation at $\epsilon_{c,1}=0.014$ can be identified as those caused by the remaining modes are negligible. The instabilities by a multitude of continuous spectrum modes are somewhat more appreciable in panel (b). Panel (c) corresponds to a point with a damping higher than $\epsilon_{c,2}=0.175$, i.e. past the bifurcation caused by the translational mode. Finally, in panel (d), there is a third quartet that emerges from the band of instabilities caused by the continuous spectrum. Notice that this quartet corresponds to the third line from the top at Fig.\ref{fig1}b.}
\label{fig1b}
\end{figure}

The instability in the present setting
stems from the collision of modes of the continuous spectrum (which are discretized in our finite domain computation), as shown in Fig. \ref{fig1}c. In
this panel, an unstable mode emerges that persists for every $\epsilon>\epsilon_{c,1}=0.014$.
Furthermore, as mentioned above, the translational mode departs from $+1$ for $\epsilon\neq 0$ and gives rise to further instabilities past $\epsilon_{c,2}=0.175$ (see Fig. \ref{fig1}a) where the translational mode bifurcates
into a quartet upon collision with a mode of the continuous spectrum. Additional computations (not shown here) have confirmed that the dominant instabilities observed persist on a larger domain and also with a smaller
discretization spacing $h$ on the same domain.
Notice also that there is a cascade of additional Hopf bifurcations caused by the collision between different ones among the extended modes at angles $\theta\neq0$, as can be seen if Fig.~\ref{fig1}c and also Fig.~\ref{fig1b}. However, these are
considerably weaker than the dominant instability,
hence they will not be considered in further detail herein.
Typical examples of the full Floquet spectrum for different values of
$\epsilon$ are given in Fig.\ref{fig1b}. It is worthwhile to note
that while it is an interesting question for further study in its own right
why the
translational mode instability occurs for $\epsilon > \epsilon_{c,2}$
(and not for smaller parameter values), it is expected that the multitude
of Hopf bifurcations occurring for smaller values of $\epsilon$ contributes
to this critical value.

\begin{figure}
\begin{tabular}{cc}
\includegraphics[width=6cm]{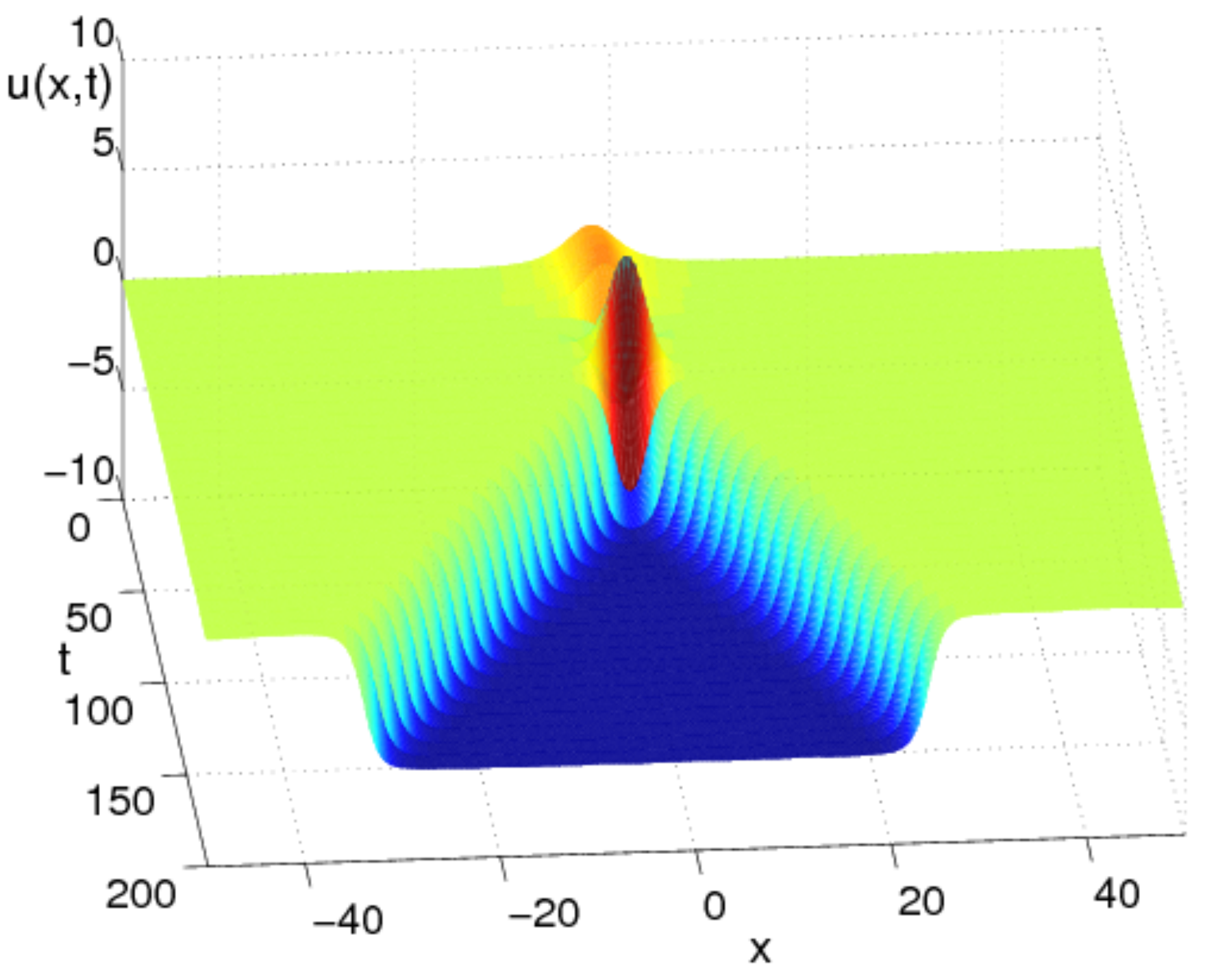} &
\includegraphics[width=6cm]{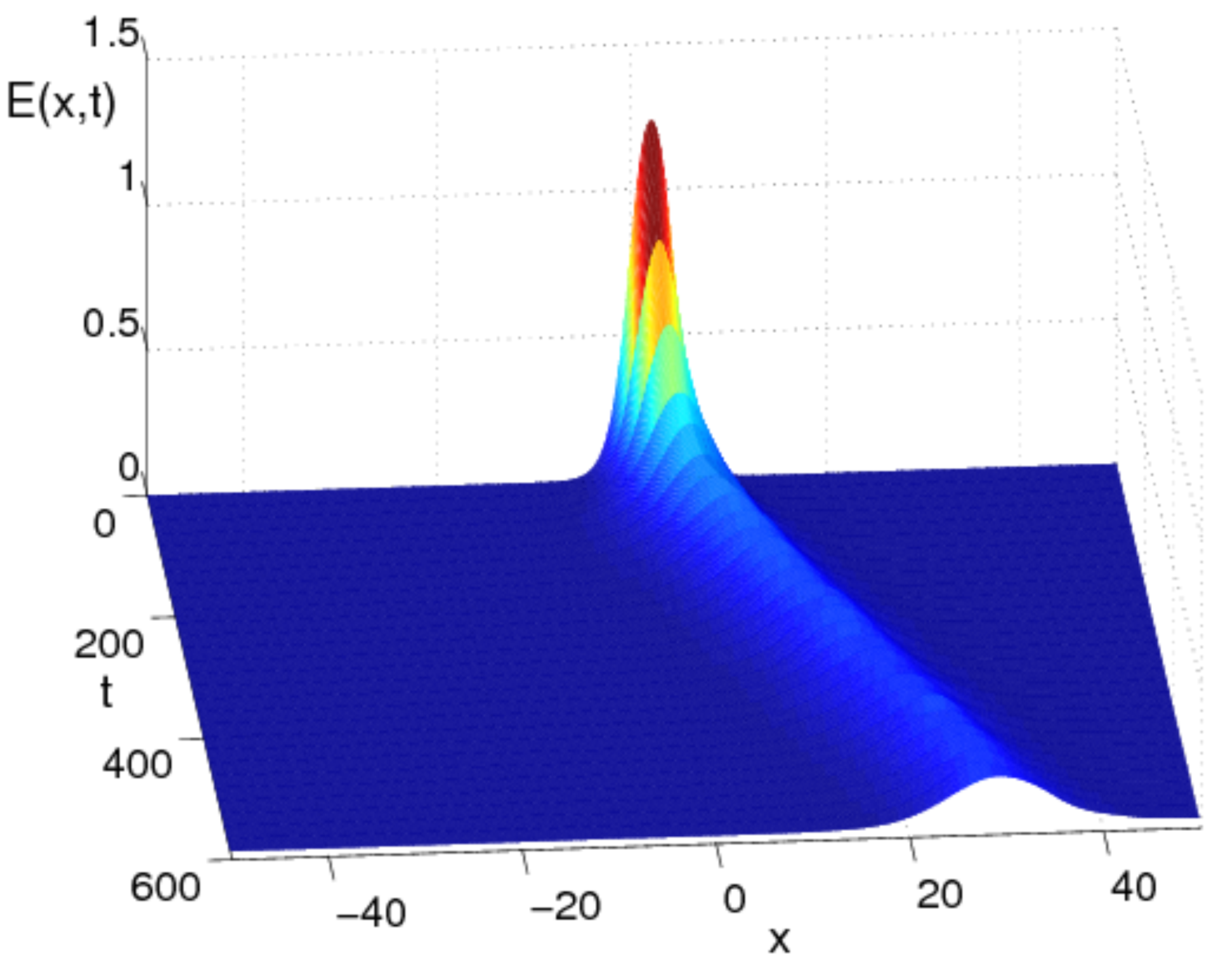} \\
\includegraphics[width=6cm]{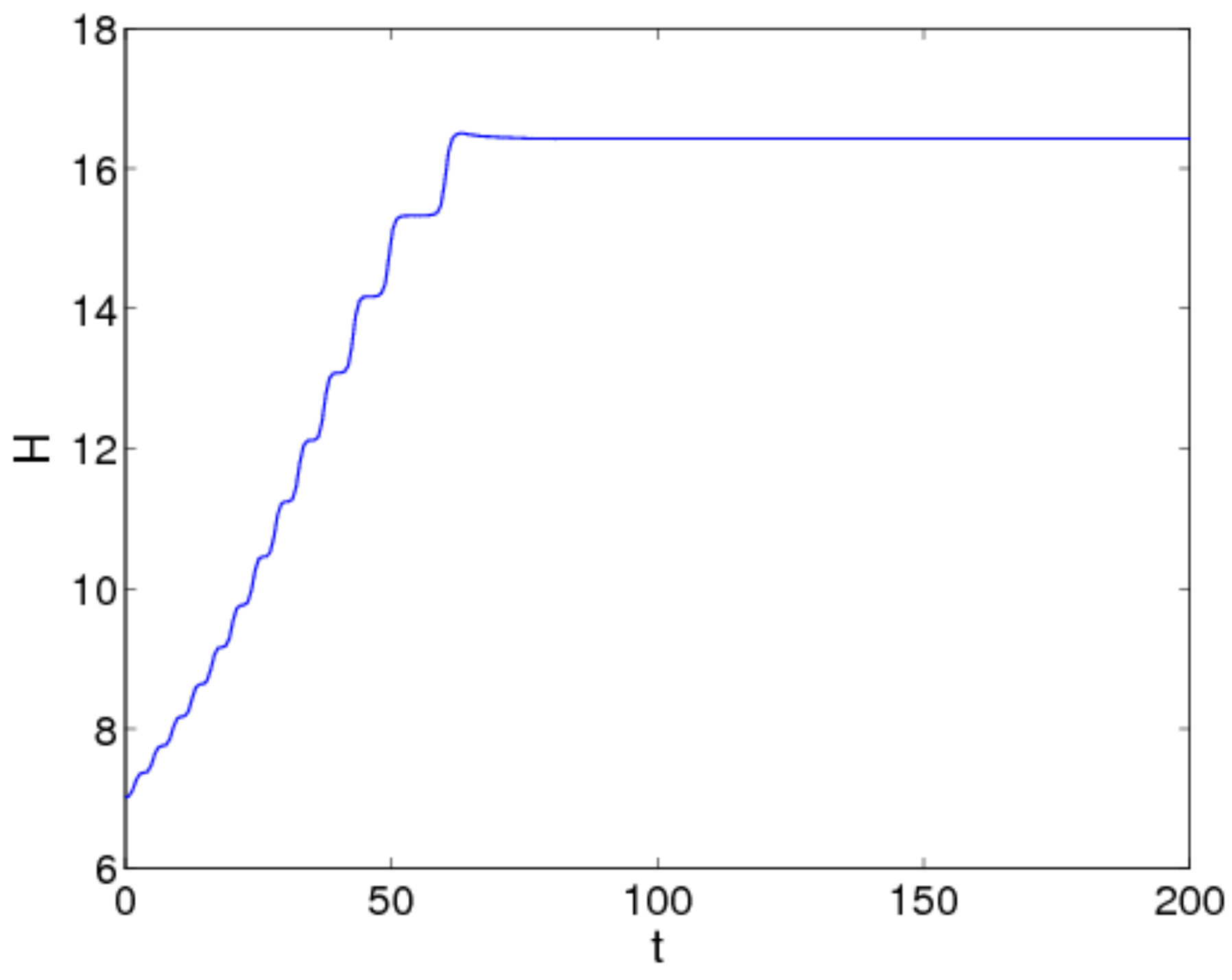} &
\includegraphics[width=6cm]{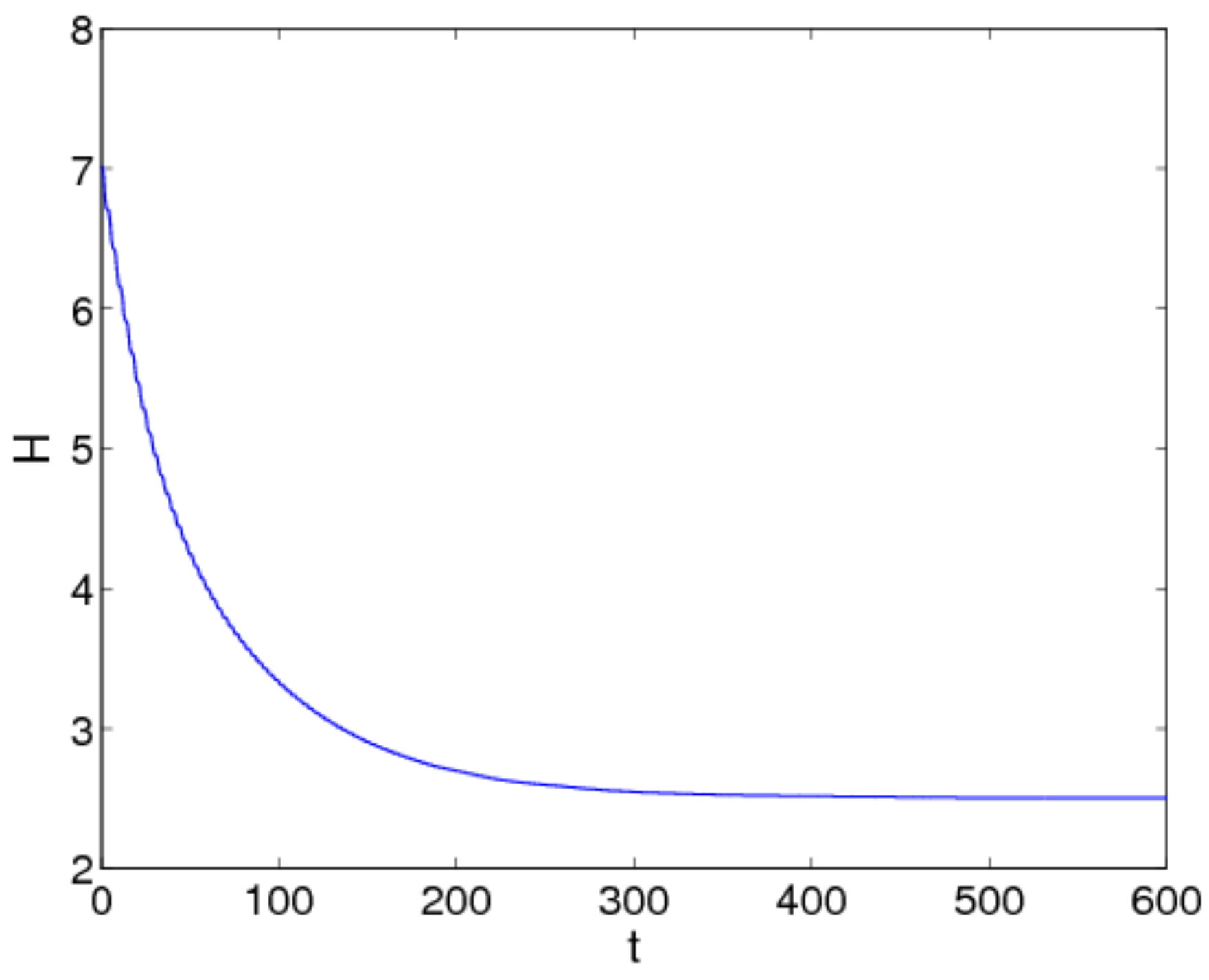} \\
\end{tabular}
\caption{An unstable breather with frequency $a=0.9$, $\mathcal{PT}$-symmetry perturbation parameter $\epsilon=0.05$ and $\gamma(x)=x\exp[-(x/2)^2]$, initially centered at $x=0$ is displaced to $x_0=-2$ (left panels) or $x_0=2$ (right panels) and subsequently used as initial condition for a dynamical simulation. In the former case, the breather is inside the gain region and transforms through its increase of energy into a kink-antikink pair. In the latter case, the breather is located in the loss region and becomes mobile, while losing energy.
The top left panel displays the evolution of the breather excitation
$u(x,t)$ whereas the top right one, for the sake of better visualization,
represents the energy density $E(x,t)$ of (\ref{eqnh}).
The bottom panels show the evolution of the Hamiltonian
energy of Eq.~(\ref{eqnh}).}
\label{fig2}
\end{figure}

\begin{figure}
\begin{tabular}{cc}
\includegraphics[width=6cm]{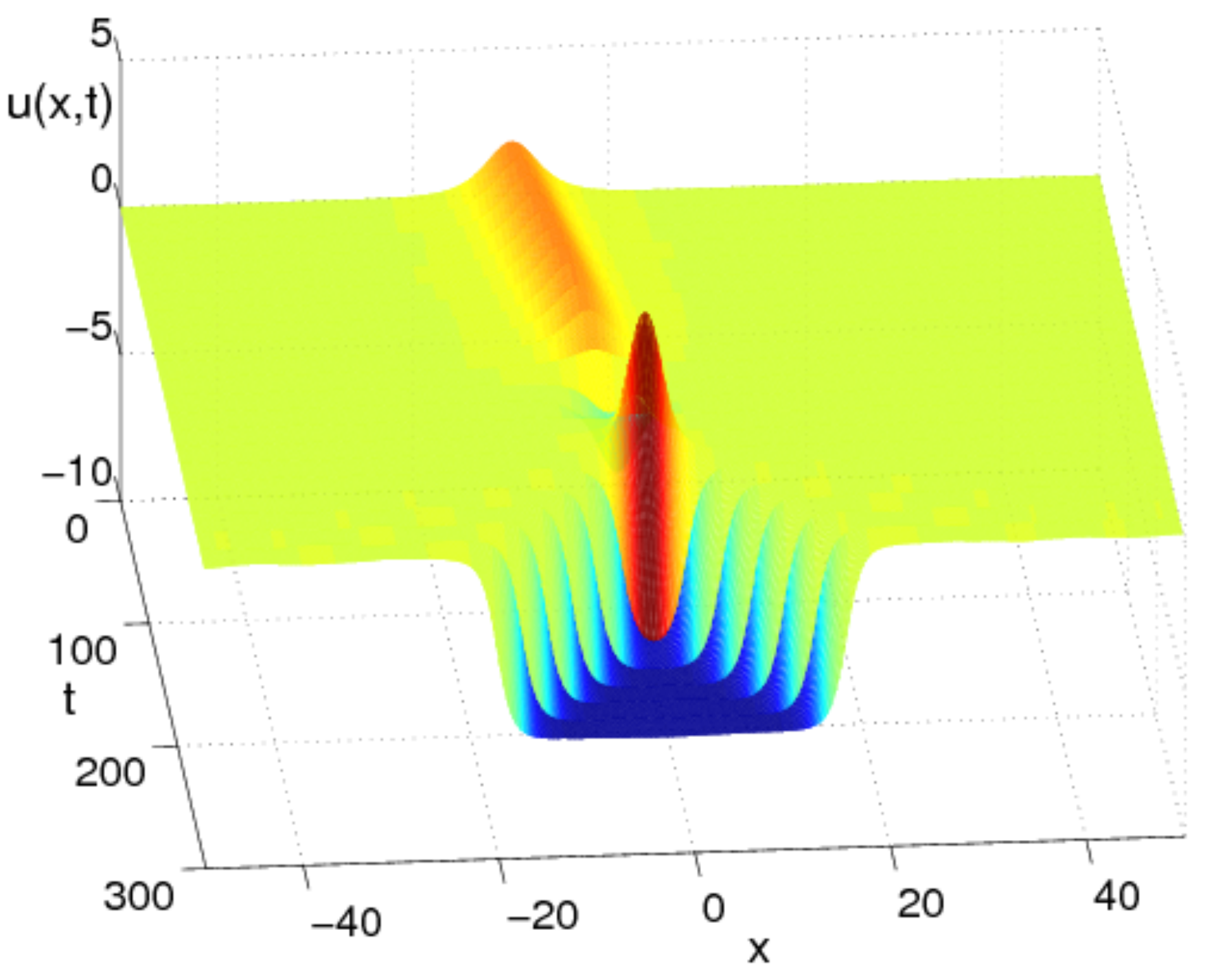} &
\includegraphics[width=6cm]{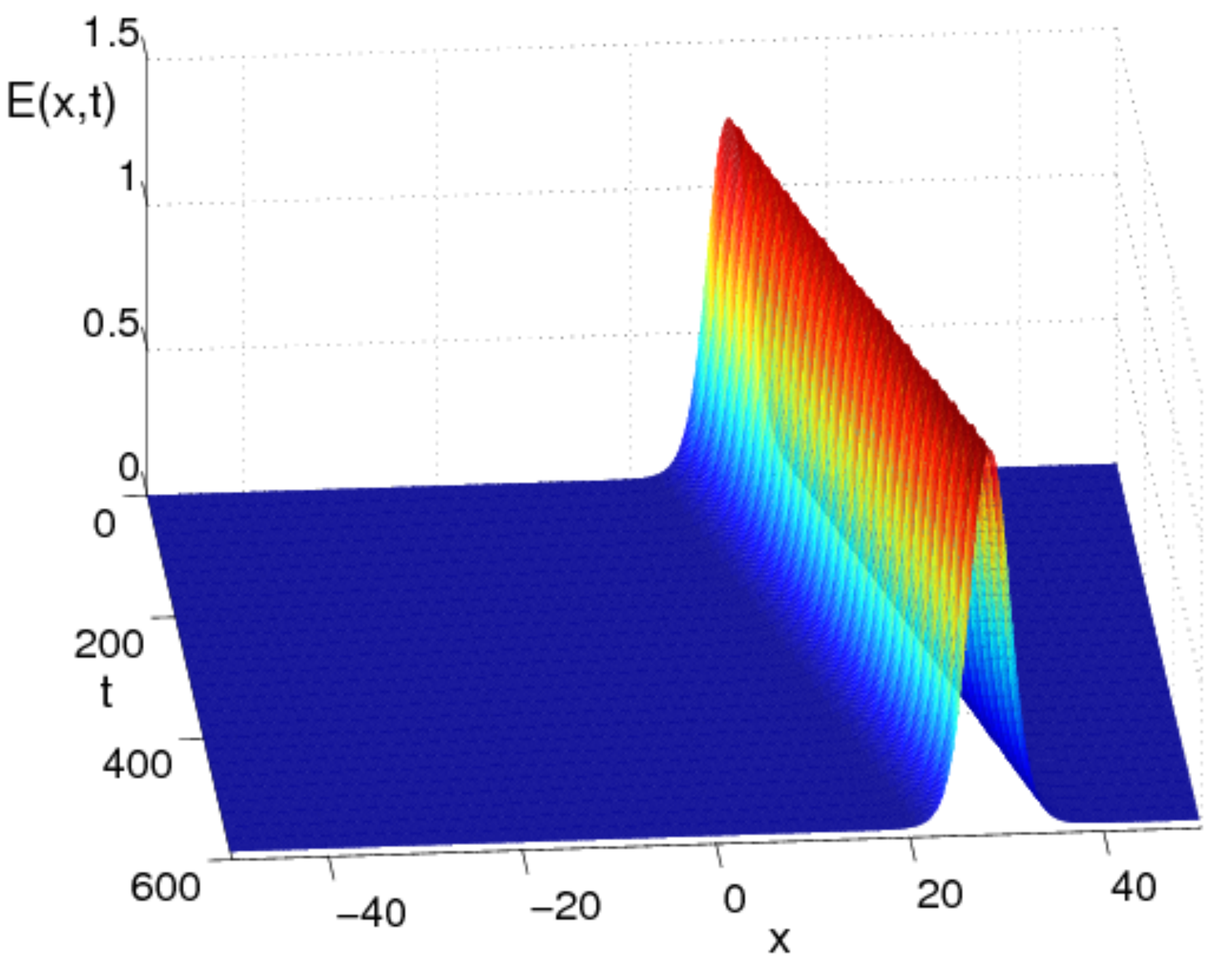} \\
\includegraphics[width=6cm]{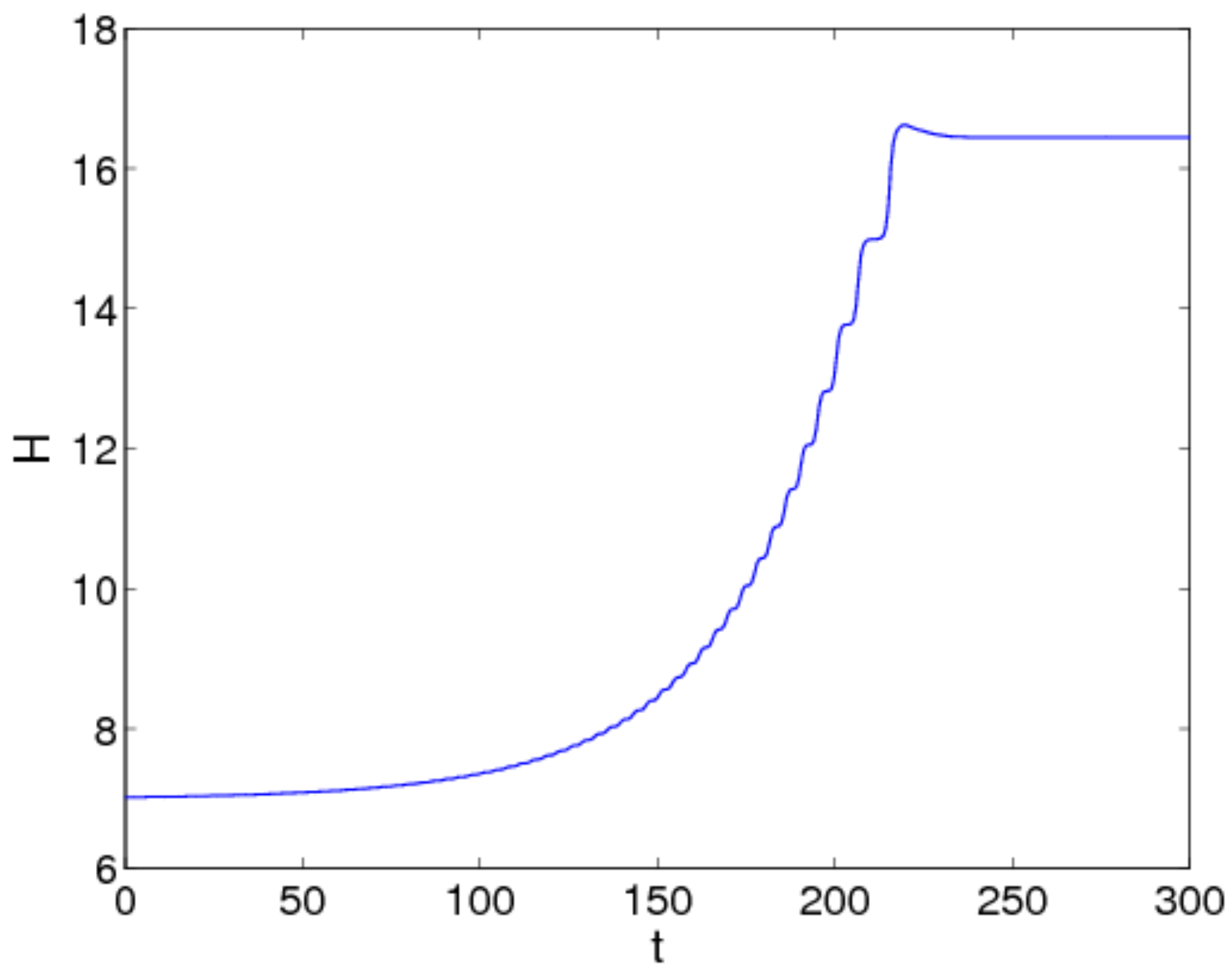} &
\includegraphics[width=6cm]{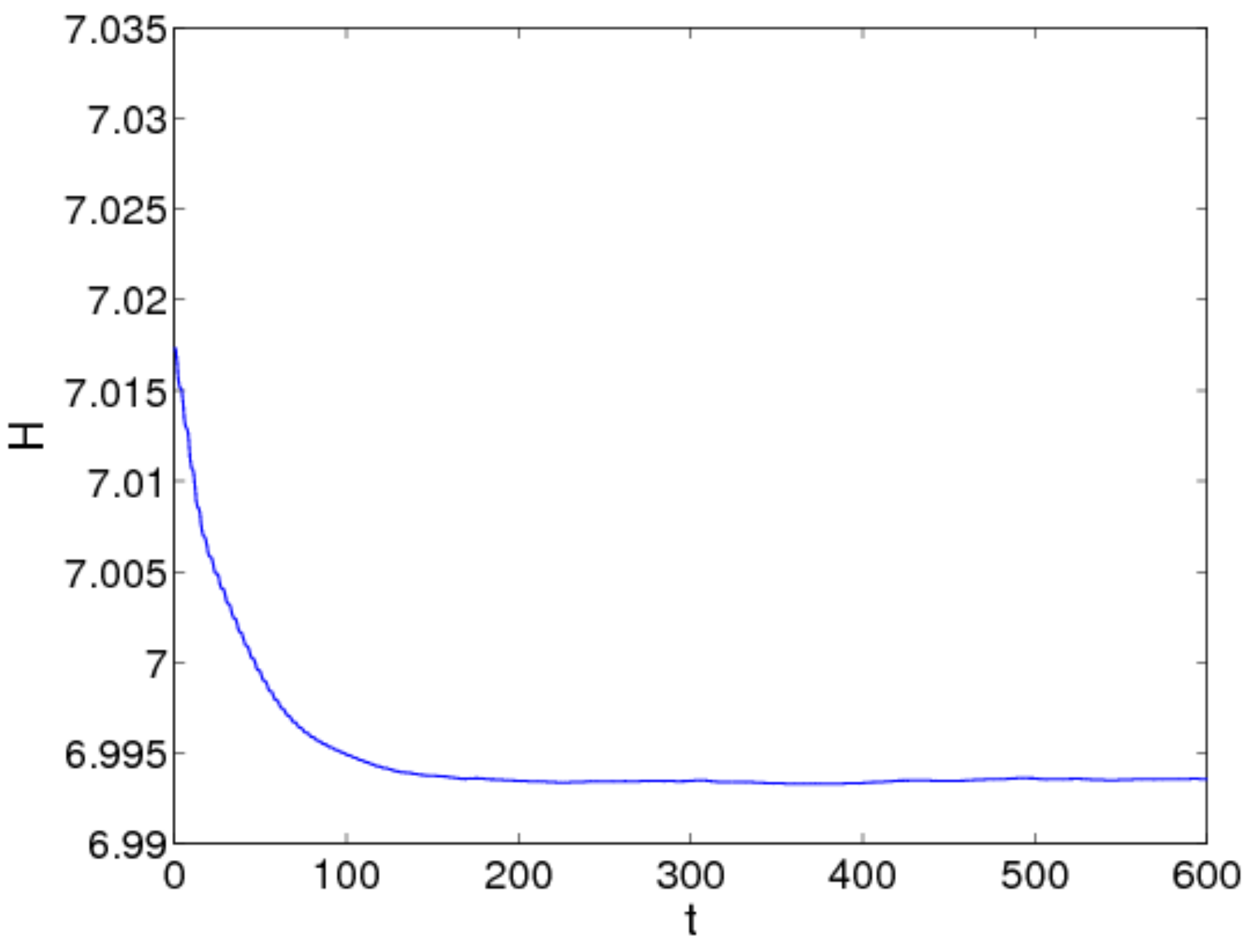} \\
\end{tabular}
\caption{Same as Fig. \ref{fig2} but for the breather displaced to $x_0=\pm10$.}
\label{fig3}
\end{figure}

We now turn to the evolutionary dynamics of the breather in this $\mathcal{P T}$-symmetric system. The most striking dynamical feature arises when the unstable breather at $x=0$ with $\epsilon\neq0$ is used as initial condition for a simulation with its center displaced to a position $x_0$. If $x_0>0$, the breather spontaneously moves in a lossy fashion
through the semi-line $x>0$; on the contrary, if $x_0 < 0$,
the breather transforms into a kink-antikink pair. Examples of this type
of dynamics are displayed in Figs. \ref{fig2} and Fig. \ref{fig3}. In the former case, the breather is displaced by a distance of $2$ towards the gain side ($x_0=-2$, left panels) and also towards the lossy side ($x_0=2$, right panels). In the former case, the bottom panels elucidate the evolution of the Hamiltonian energy functional of the form

\begin{eqnarray}
H = \int E(x,t) \mathrm{d}x, \qquad \text{with}\ E(x,t)=\frac{u_t^2}{2} + \frac{u_x^2}{2} + \left(1- \cos(u) \right);
\label{eqnh}
\end{eqnarray}
i.e., $E(x,t)$ represents the energy density at a given space point $x$
(and a given time $t$).

When the breather is centered on the gain side, we observe that its Hamiltonian energy grows from its initial value until it hits the ``nucleation threshold'' of $H=16$, at which time the structure can transform itself into a kink-antikink pair, given that that is the energy of such a pair (the individual
energy of the kink and of the antikink  is $H=8$)~\cite{dodd}. On the lossy side, on the other hand, the energy is observed to continuously decrease leading to the eventual
``annihilation'' of the breather. Figure~\ref{fig3} essentially illustrates that the above phenomenology is generic along the gain and loss sides of our $\gamma(x)$, although a much larger choice of $x_0$ ($= \pm 10$ in the latter case) decreases the rate of emergence of
the relevant phenomenology, as the gain and loss are considerably weaker
at that location.

Although as the initial profile for the simulations we have made use of
that of the unstable stationary breathers, the above mentioned scenarios would be the same if stable stationary breathers were used as initial condition.

\begin{figure}
\begin{tabular}{cc}
\includegraphics[width=6cm]{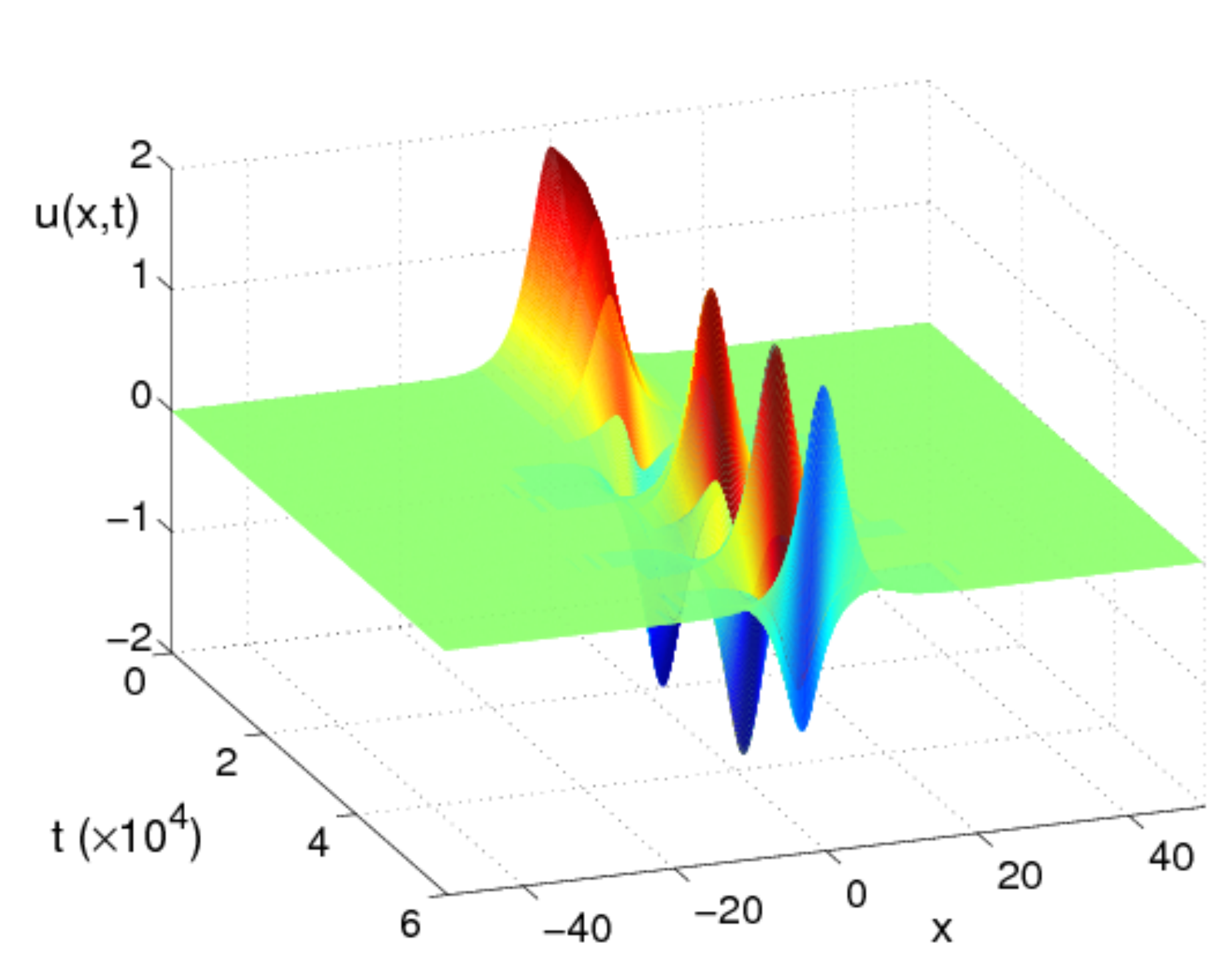} &
\includegraphics[width=6cm]{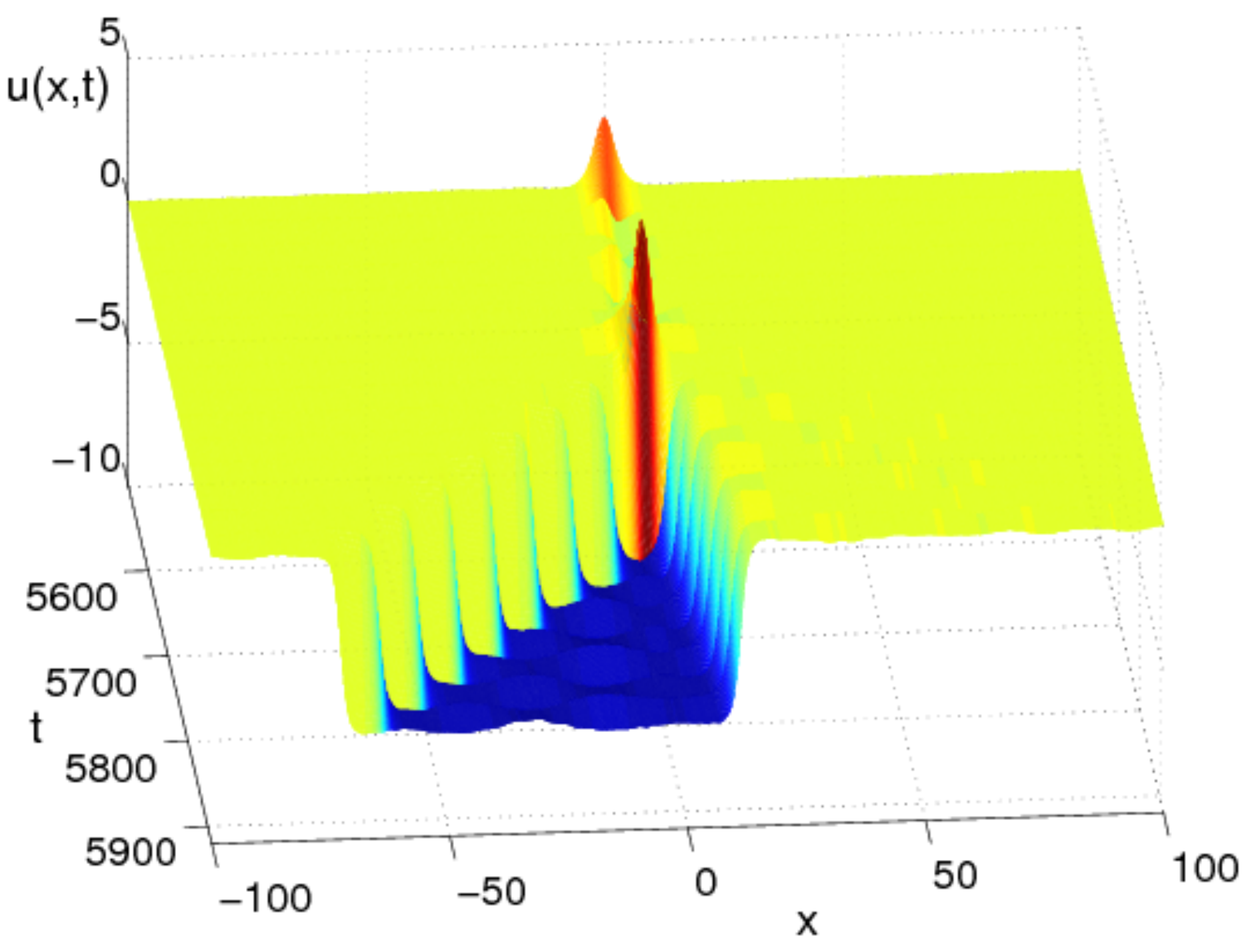} \\
\includegraphics[width=6cm]{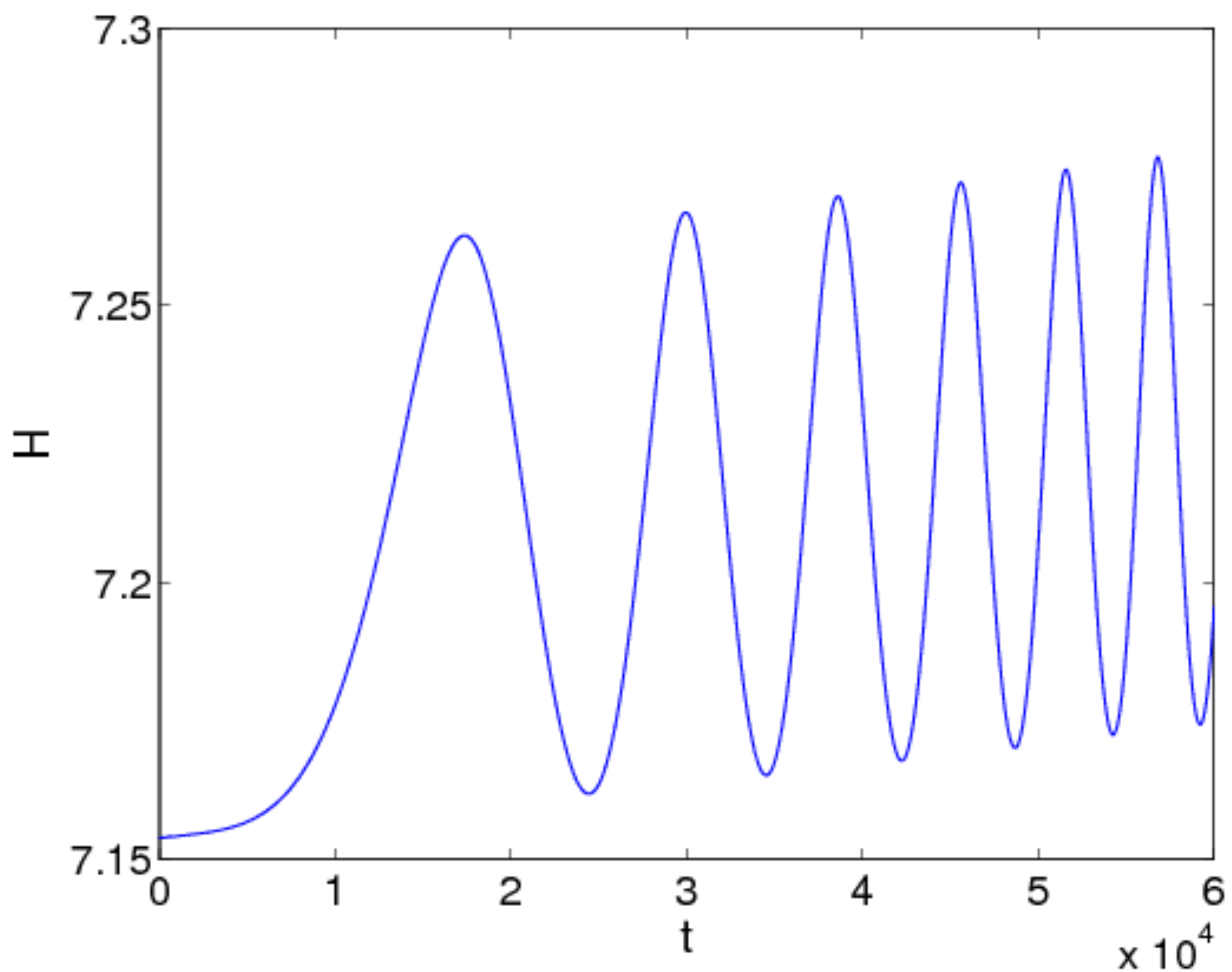} &
\includegraphics[width=6cm]{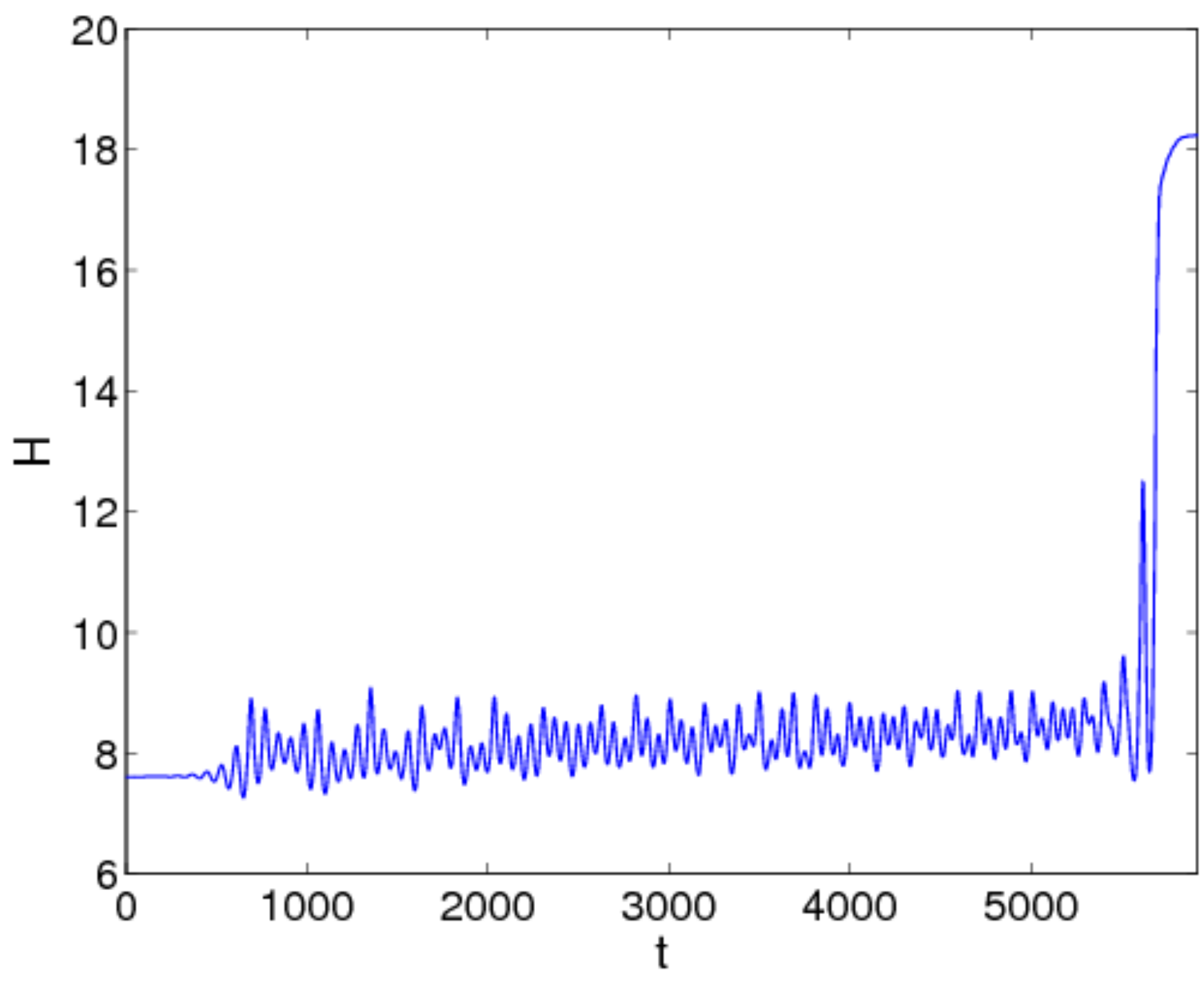} \\
\end{tabular}
\caption{Dynamical evolution of an unstable breather with $\epsilon=0.1$ (left) and $\epsilon=0.18$ (right) located at $x_0=0$. Notice that in this figure and the previous ones, the points are taken with a rough time discretization that does not allow to observe the internal oscillation frequency $a$.}
\label{fig4}
\end{figure}

We have finally considered the evolution of unstable breathers located at $x_0=0$. If the breather is in the interval $\epsilon_{c,1}<\epsilon<\epsilon_{c,2}$, i.e. after the first bifurcation, the breather becomes quasi-periodic. For $\epsilon>\epsilon_{c,2}$ the breather transforms into a kink-antikink pair, similarly to the case of a breather located at the gain region. In both cases, the eigenvectors associated with the instabilities are localized in the gain ($x<0$) side. Fig. \ref{fig4} shows two prototypical examples of these kinds of evolution.
It is interesting to point out in this context that for this
larger value of $\epsilon$, it can be seen that upon nucleation of
the kink-antikink pair, the structure emerging on the gain side propagates
unhindered, while the one appearing on the lossy side seems to
decelerate and nearly stop. This is in line with earlier numerical
(and semi-analytical) observations in Klein-Gordon models; cf.
Ref.~\cite{galley}.

\section{Conclusions \& Future Challenges}

In the present work, we have considered a prototypical example
of the effect of $\mathcal{P T}$-symmetry on continuum breathers
within the realm of the sine-Gordon equation.
It has been shown through a Melnikov type calculation
 that the breathers are especially ``delicate''
persisting only at the special location of the interface
between the gain and the loss. This non-robustness apparently
renders these breathers linearly unstable, through a Hopf bifurcation
as has been revealed in our Floquet analysis of the linearization
problem and its monodromy matrix. Finally, the nonlinear dynamical
evolution of the breathers has shown that when tilted towards the
gain side, their energy grows until it is sufficient to nucleate
a pair of a kink and anti-kink that subsequently separate from
each other. On the other hand, in the case of the lossy side,
it can be seen that the breathers gradually lose their energy,
eventually being dissipated away.

Naturally many extensions of the present work can be considered.
Perhaps the most natural one would be to explore the realm of
discrete systems where breathers instead of being rather special,
they are fairly generic~\cite{macaub} (under suitable non-resonance
conditions). Understanding their persistence in such discrete settings
might, in turn, reveal their potential observability in experimental
settings (such as electrical lattices). Additionally, while here
we have considered the special case of an exponentially decaying
$\gamma(x)$ which corresponds purely to gain for $x<0$ and loss
for $x>0$, it would be quite relevant to explore more complex
forms of $\gamma(x)$, possibly involving oscillations  of gain
and loss (within the span of the breather) which  may
possess more complex breather existence and stability properties.
Such studies are currently in progress and will be reported in future
publications.

\section*{Appendix: Existence and stability of breathers}

In order to perform a numerical analysis of the existence and stability of discrete breathers of frequency $a$ (which is in this case the natural parameter of the breather and assumes values $0<a<1$), we firstly need to discretize the spatial partial derivative in the sine-Gordon equation. We consider a finite-difference scheme so that

\begin{equation}
    u_{xx}\approx \frac{1}{h^2}(u_{n+1}+u_{n-1}-2u_n)
\end{equation}
with $h$ being the discretization parameter. This transforms equation (\ref{dyn}) into a set of $N$ coupled differential equations:

\begin{eqnarray}\label{dyndisc}
    \ddot u_n  +\epsilon \gamma(nh) \dot u_n + \sin u_n + \frac{1}{h^2}(u_{n+1}+u_{n-1}-2u_n)= 0
\end{eqnarray}

In order to calculate breathers in the $\cP\cT$-symmetric sine-Gordon model, we make use of a Fourier space implementation of the discretized dynamical equations (\ref{dyndisc}) and continuations in frequency or gain/loss parameter are performed via a path-following (Newton-Raphson) method. Fourier space methods are based on the fact that the solutions are $T$-periodic, with $T=2\pi/a$; for a detailed explanation of these methods, the reader is referred to Refs.~\cite{AMM99,Marin,Cuevas}. The method has the advantage, among others, of providing an explicit, analytical form of the Jacobian. Thus, the solution for the discretized system can be expressed in terms of a truncated Fourier series expansion:

\begin{equation}\label{series}
    u_n(t)=\sum_{k=-k_m}^{k_m} z_{k,n}\exp(\ii k a t)\ ,
\end{equation}
with $k_m$ being the maximum of the absolute value of the running index $k$ in our Galerkin truncation of the full Fourier series solution. In the numerics, $k_m$ has been chosen as 15. After the introduction of (\ref{series}), the equations (\ref{dyndisc}) yield a set of $N\times(2k_m+1)$ nonlinear, coupled algebraic equations:

\begin{equation}\label{Fourier}
    F_{k,n} \equiv -a^2k^2z_{k,n}-\ii\gamma(nh)a k z_{k,n}+{\cal F}_{k,n}-\frac{1}{h^2}(z_{k,n+1}+z_{k,n-1}-2z_{k,n})=0\ .
\end{equation}

Here, ${\cal F}_{k,n}$ denotes the Discrete Fourier Transform:

\begin{equation}
    {\cal F}_{k,n}=\frac{1}{\cal N}\sum_{q=-k_m}^{k_m}\sin\left(\sum_{p=-k_m}^{k_m}z_p\exp\left[\ii \frac{2\pi p n}{\cal N}\right]\right)
    \exp\left[-\ii \frac{2\pi k q}{\cal N}\right],
\end{equation}
with ${\cal N}=2k_m+1$. As $u_n(t)$ must be a real function, it implies that $z_{-k,n}=z^*_{k,n}$.

In order to study the spectral stability of periodic orbits, we introduce a small perturbation $\xi_n$ to a given solution $u_{n,0}$ of Eq. (\ref{dyndisc}) according to $u_n=u_{n,0}+\xi_n$. Then, the equations satisfied to first order in $\xi_n$ read:

\begin{equation}\label{stab}
    \ddot\xi_n+\cos(u_0)\xi_n-\gamma(nh) \dot\xi_n-\frac{1}{h^2}(\xi_{n+1}+\xi_{n-1}-2\xi_n) = 0\ .
\end{equation}

In order to study the spectral (linear) stability analysis of the relevant solution, a Floquet analysis can be performed if there exists $T\in\mathbb{R}$ so that the map $\{u_n(0)\}\rightarrow \{u_n(T)\}$ has a fixed point (which constitutes a periodic orbit of the original system). Then, the stability
properties are given by the spectrum of the Floquet operator $\mathcal{M}$ (whose matrix representation is the monodromy) defined as:

\begin{equation}\label{eq:monodromy}
    \left(\begin{array}{c} \{\xi_{n}(T)\} \\ \{\dot\xi_{n}(T)\} \\ \end{array}
    \right)=\mathcal{M}\left(\begin{array}{c} \{\xi_{n}(0)\} \\ \{\dot\xi_{n}(0)\} \\ \end{array}
    \right) .
\end{equation}

The $2N\times2N$ monodromy eigenvalues $\Lambda=\exp(\ii\theta)$ are dubbed the {\em Floquet multipliers} and $\theta$ are denoted as {\em Floquet exponents}
(FEs). This operator is real, which implies that there is always a pair of multipliers at $1$ (corresponding to the so-called phase and growth modes~\cite{Marin,Cuevas}) and that the eigenvalues come in pairs $\{\Lambda,\Lambda^*\}$.

\section*{Acknowledgments}

We are indebted to Ricardo Carretero-Gonz\'alez for his technical support.
P.G.K. also acknowledges support from the National Science Foundation under grants CMMI-1000337,
DMS-1312856, from the Binational Science Foundation under grant 2010239, from FP7-People under grant IRSES-
606096 and from the US-AFOSR under grant FA9550-12-10332.


\begin{thebibliography}{99}

\bibitem{bend1} C.M. Bender and S. Boettcher, Phys. Rev. Lett. {\bf 80}, 5243 (1998).

\bibitem{bend2} C.M. Bender, S. Boettcher, and P.N. Meisinger, J. Math. Phys. {\bf 40}, 2201 (1999)

\bibitem{bend3} C. M. Bender,
Rep.\ Prog.\ Phys.\ {\bf 70} (2007) 947--1018.

\bibitem{Muga} A. Ruschhaupt, F. Delgado, and J.G. Muga,
J. Phys. A: Math. Gen.
{\bf 38} (2005) L171--L176.

\bibitem{ziad} Z. H. Musslimani, K. G. Makris, R. El-Ganainy, and D. N. Christodoulides,
Phys. Rev. Lett. {\bf 100} (2008) 030402 (4 pages).

\bibitem{Ramezani} H. Ramezani, T. Kottos, R. El-Ganainy, and
D.N. Christodoulides,
Phys. Rev. A {\bf 82} (2010), 043803 (6 pages).

\bibitem{salamo} A. Guo, G.J. Salamo, D. Duchesne, R. Morandotti,
M. Volatier-Ravat, V.
Aimez, G. A. Siviloglou, and D. N. Christodoulides,
Phys. Rev. Lett. {\bf 103},
093902 (2009).

\bibitem{dncnat}
C.E. R{\"u}ter, K.G. Makris, R. El-Ganainy, D.N. Christodoulides, M.Segev, and D. Kip,
Nature Physics {\bf 6} (2010) 192--195.

\bibitem{whisper} B. Peng, S.K. Ozdemir, F. Lei, F. Monifi, M. Gianfreda,
G.L. Long, S. Fan, F. Nori, C.M. Bender, L. Yang, Nature Physics {\bf 10} (2014) 394–-398.

\bibitem{bend_mech} C.M. Bender, B.J. Berntson, D. Parker and
E. Samuel,
Am. J. Phys. {\bf 81}, 173 (2013).

\bibitem{R21} J. Schindler, A. Li, M. C. Zheng, F. M. Ellis, and T. Kottos,
Phys. Rev. A {\bf 84}, 040101 (2011).

\bibitem{tsampas2} H. Ramezani, J. Schindler, F. M. Ellis, U. G{\"u}nther,
and T. Kottos,
Phys. Rev. A {\bf 85}, 062122 (2012).

\bibitem{Abdullaev} F.Kh. Abdullaev, Y.V. Kartashov, V.V. Konotop, and D.A. Zezyulin,
Phys. Rev. A {\bf 83}, 041805(R) (2011).

\bibitem{baras2} N. V. Alexeeva, I. V. Barashenkov, A.A. Sukhorukov,
and Yu.S. Kivshar,
Phys. Rev. A {\bf 85}, 063837 (2012)

\bibitem{baras1} I.V. Barashenkov, S.V. Suchkov, A.A. Sukhorukov,
S.V. Dmitriev and Yu.S. Kivshar,
Phys. Rev. A {\bf 86}, 053809 (2012)

\bibitem{malom1} R. Driben and B.A. Malomed,
Opt. Lett. {\bf 36}, 4323 (2011).

\bibitem{malom2} R. Driben and B.A. Malomed,
EPL {\bf 96}, 51001 (2011).

\bibitem{Nixon} S. Nixon, L. Ge, and J. Yang,
Phys. Rev. A {\bf 85}, 023822 (2012).

\bibitem{Dmitriev} S.V. Dmitriev, A.A. Sukhorukov, and Yu.S. Kivshar,
Opt. Lett.
{\bf 35}, 2976 (2010).

\bibitem{Pelin1} V.V. Konotop, D.E. Pelinovsky, and D.A. Zezyulin,
EPL {\bf 100}, 56006 (2012).

\bibitem{Sukh}  A.A. Sukhorukov, S.V. Dmitriev, S.V. Suchkov, and Yu.S. Kivshar,
Opt. Lett. {\bf 37}, 2148 (2012).

\bibitem{bludov} Yu.V. Bludov, R. Driben, V.V. Konotop, B.A. Malomed,
J. Opt. {\bf 15}, 064010 (2013).

\bibitem{yang2} J. Yang, Phys. Lett. A {\bf 378}, 367--373 (2014)
and also J. Yang, Opt. Lett. {\bf 39}, 1133-1136 (2014).


\bibitem{Li} K. Li and P.G. Kevrekidis,
Phys. Rev. E {\bf 83}, 066608 (2011).

\bibitem{Guenter} K. Li, P.G. Kevrekidis, B.A. Malomed, and U. G\"{u}nther,
J. Phys. A Math. Theor. {\bf 45}, 444021 (2012)

\bibitem{suchkov} S.V. Suchkov, B.A. Malomed, S.V. Dmitriev and
Yu.S. Kivshar,
Phys. Rev. E {\bf 84}, 046609 (2011).

\bibitem{Sukhorukov} A.A. Sukhorukov, Z. Xu, and  Yu.S. Kivshar,
Phys. Rev. A {\bf 82}, 043818 (2010).

\bibitem{ZK} D.A. Zezyulin and V.V. Konotop,
Phys. Rev. Lett. {\bf 108}, 213906 (2012).

\bibitem{Flachbar} N. V. Alexeeva, I. V. Barashenkov, K. Rayanov, S. Flach,
 arXiv:1308.5862.

\bibitem{Flachbar2} I. V. Barashenkov, G.S Jackson, S. Flach,
Phys. Rev. A {\bf 88}, 053817 (2013).

\bibitem{tsironis} N. Lazarides and G. P. Tsironis,
Phys. Rev. Lett. 110, 053901 (2013).

\bibitem{ptkg} J. Cuevas, P.G. Kevrekidis, A. Saxena and A. Khare,
Phys. Rev. A {\bf 88}, 032108 (2013).

\bibitem{demirkaya} A. Demirkaya,
D. J. Frantzeskakis, P. G. Kevrekidis A. Saxena, and A. Stefanov,
Phys. Rev. E {\bf 88}, 023203 (2013).

\bibitem{demirkaya2} A. Demirkaya, M. Stanislavova, A. Stefanov,
T. Kapitula and P.G. Kevrekidis,
arXiv:1402.2942.

\bibitem{gianfreda} I.V. Barashenkov and M. Gianfreda,
J. Phys. A: Math. Theor. {\bf 47} 282001 (2014).

\bibitem{factor} N. Bender, S. Factor, J. D. Bodyfelt, H. Ramezani, D. N. Christodoulides, F. M. Ellis, and T. Kottos, Phys. Rev. Lett. {\bf 110}, 234101
(2013).

\bibitem{birnir} B. Birnir, H.P. McKean and A. Weinstein,
Comm. Pure Appl. Math. {\bf 47}, 1043 (1994).

\bibitem{denzler} J. Denzler,
Comm. Math. Phys. {\bf 158}, 397 (1993).

\bibitem{shatah} J. Shatah, C.C. Zeng,
Nonlinearity {\bf 16}, 591 (2003).

\bibitem{lu} N. Lu,
J. Differ. Equations {\bf 256}, 745  (2014).

\bibitem{guggen} J. Guggenheimer, P. Holmes,
{\it Nonlinear Oscillations, Dynamical Systems and Bifurcation
of Vector Fields}, Springer-Verlag (New York, 1983).

\bibitem{Aubry} S. Aubry,
Physica D {\bf 103}, 201 (1997).

\bibitem{flachgor} S. Flach and A. V. Gorbach, Phys. Rep. {\bf 467}, 1 (2008).


\bibitem{AMM99}
J.F.R. Archilla, R.S. MacKay, and J.L. Mar\'{\i}n,
\newblock {\em Physica} D {\bf 134}, 406 (1999).

\bibitem{Marin}
J.L. Mar\'{\i}n,
\newblock {I}ntrinsic {L}ocalized {M}odes in nonlinear lattices.
\newblock {\em Ph{D} {T}hesis}, University of {Z}aragoza (1999).

\bibitem{Cuevas}
J. Cuevas
\newblock Localization and energy transfer in anharmonic inhomogeneus lattices
\newblock {\em Ph{D} {T}hesis}, University of Sevilla (2003).


 \bibitem{dodd} R.K. Dodd, J.C. Eilbeck, J.D. Gibbon and
H.C. Morris,
{\it Solitons and Nonlinear Wave Equations}, Academic Press (London, 1982).

\bibitem{galley} P.G. Kevrekidis,
Phys. Rev. A {\bf 89}, 010102(R) (2014).



\bibitem{macaub}  R.S. MacKay and S. Aubry,
Nonlinearity {\bf 7}, 1623 (1994).


\end{thebibliography}
\end{document}